\documentclass[12pt,twoside,english]{extarticle}
\usepackage{lmodern}
\usepackage{helvet}
\usepackage[T1]{fontenc}
\usepackage[latin9]{inputenc}
\usepackage{geometry}
\usepackage{color}
\usepackage{babel}
\usepackage{float}
\usepackage{mathrsfs}
\usepackage{amsmath}
\usepackage{amsthm}
\usepackage{amssymb}
\usepackage{graphicx}
\usepackage{setspace}
\usepackage[unicode=true,pdfusetitle,
 bookmarks=true,bookmarksnumbered=false,bookmarksopen=false,
 breaklinks=true,pdfborder={0 0 0},pdfborderstyle={},backref=false,colorlinks=true]
 {hyperref}

\makeatletter

\providecommand{\tabularnewline}{\\}

\numberwithin{equation}{section}
\numberwithin{table}{section}
\numberwithin{figure}{section}
\usepackage[natbibapa]{apacite}
\theoremstyle{plain}
\newtheorem{assumption}{\protect\assumptionname}
\theoremstyle{plain}
\newtheorem{prop}{\protect\propositionname}[section]
\theoremstyle{definition}
\newtheorem{defn}{\protect\definitionname}[section]
\theoremstyle{plain}
\newtheorem{lem}{\protect\lemmaname}[section]
\theoremstyle{definition}
\newtheorem{condition}{\protect\conditionname}
\theoremstyle{plain}
\newtheorem{thm}{\protect\theoremname}[section]
\theoremstyle{plain}
\newtheorem{cor}{\protect\corollaryname}[section]
\theoremstyle{plain}
\newtheorem{lyxalgorithm}{\protect\algorithmname}

\usepackage{graphicx}
\usepackage{amsmath}
\usepackage{dcolumn}
\usepackage{listings}
\usepackage{color}
\usepackage{setspace}
\usepackage[normalem]{ulem}  
\definecolor{hellgelb}{rgb}{1,1,0.8}
\definecolor{colKeys}{rgb}{0,0,1}
\definecolor{colIdentifier}{rgb}{0,0,0}
\definecolor{colComments}{rgb}{1,0,0}
\definecolor{colString}{rgb}{0,0.5,0}

\usepackage{lmodern}
\usepackage[T1]{fontenc}
\usepackage{geometry}
\usepackage{fancyhdr}
\usepackage{amsthm}
\usepackage{thmtools}
\usepackage{amsmath}
\usepackage{amssymb}
\usepackage{esint}
\usepackage{multirow}
\usepackage{mathrsfs} 
\usepackage{textgreek}
\usepackage{amsfonts}
\usepackage{amssymb}
\usepackage{mathrsfs}

\usepackage[USenglish]{isodate}
\usepackage{chngcntr}

\numberwithin{equation}{section}
\numberwithin{table}{section}
\numberwithin{assumption}{section}
\counterwithout{figure}{section}
\counterwithout{table}{section}

\makeatother

  \providecommand{\algorithmname}{Algorithm}
  \providecommand{\assumptionname}{Assumption}

  \providecommand{\corollaryname}{Corollary}
  
  \providecommand{\definitionname}{Definition}

  \providecommand{\lemmaname}{Lemma}

  \providecommand{\propositionname}{Proposition}

  \providecommand{\theoremname}{Theorem}
 \providecommand{\corollaryname}{Corollary}
 \providecommand{\theoremname}{Theorem}
 
\usepackage{thmtools}

\newtheoremstyle{MyTheoremstyle}
  {\topsep} 
  {\topsep} 
  {} 
  {} 
  {\bfseries} 
  {.} 
  {.90em} 
  {} 
\theoremstyle{MyTheoremstyle} 
\theoremstyle{MyTheoremstyle} 
\theoremstyle{MyTheoremstyle} 
\theoremstyle{MyTheoremstyle} 
\theoremstyle{MyTheoremstyle}

\declaretheoremstyle[
    headfont=\bfseries,
    notefont=\normalfont,
    bodyfont=\itshape,
    headpunct=\newline,
    headformat={%
        \makebox{\NAME\ \NUMBER\ }{\NOTE}%
    },
]{theorem}

\newlength{\spacelength}
\declaretheoremstyle[
    headfont=\bfseries,
    notefont=\normalfont,
    bodyfont=\itshape,
    headpunct=\newline,
    headformat={%
        \makebox[0pt][l]{\NAME\ \NUMBER\ }\hskip-\spacelength{\NOTE}%
    },
]{theore}

 \usepackage{fancyhdr}
\usepackage{booktabs}
\usepackage{float}
\usepackage{graphicx}
\usepackage{epstopdf}
\usepackage{morefloats}
\usepackage[referable]{threeparttablex}
\usepackage{footnote}

\usepackage{setspace} 
\usepackage{graphicx,color}

\usepackage{listings}
\RequirePackage[T1]{fontenc}
\RequirePackage{ae,fancyvrb}
\DefineVerbatimEnvironment{Sinput}{Verbatim}{fontshape=sl}
\DefineVerbatimEnvironment{Soutput}{Verbatim}{}
\DefineVerbatimEnvironment{Scode}{Verbatim}{fontshape=sl}

\title{\bf Continuous Record Asymptotic Framework for Inference in Strucutral Change Models}
\author{
\textsc{\textcolor{MyBlue}{Alessandro Casini}}\thanks{Corresponding author at: Department of Economics and Finance, University of Rome Tor Vergata, Via Columbia 2, Rome, 00133, IT. 
Email: 
\texttt{\textcolor{MyBlue}{{alessandro.casini@uniroma2.it}}}.} 
\\
\small{University of Rome Tor Vergata}
\and
\textsc{\textcolor{MyBlue}{Pierre Perron}}\thanks{Department of Economics, Boston University, 270 Bay State Road, Boston, MA 02215, US. 
Email: 
\texttt{\textcolor{MyBlue}{\mbox{perron@bu.edu}}}.} 
\\
\small{Boston University}
}

\date{\small{\today} \\} 

 \makeatletter

\numberwithin{equation}{section}
\makeatother

\usepackage{babel}
\addto\captionsenglish{%
}

\usepackage{color}
\usepackage{xcolor}

\definecolor{MyRed}{rgb}{0.8,0,0}
\definecolor{MyBlue}{rgb}{0,0,0.7}
\definecolor{Green}{rgb}{0,0.5,0}
\definecolor{hellgelb}{rgb}{1,1,0.8}
\definecolor{colKeys}{rgb}{0,0,1}
\definecolor{colIdentifier}{rgb}{0,0,0}
\definecolor{colComments}{rgb}{1,0,0}
\definecolor{colString}{rgb}{0,0.5,0}

\definecolor{MyLightRed}{rgb}{2.2,0.2,0.4} 
\definecolor{MyLightRed2}{rgb}{0.6,0.2,0.3} 
\definecolor{MyLightRed2temp}{rgb}{0.6,0.2,0.3}
\definecolor{MyLightRed3}{rgb}{0.8,0.1,0.1} 
\definecolor{MyRed}{rgb}{0.7,0.0,0}

\definecolor{MyLigthBlue13}{rgb}{0,0.2,0.7}
 \definecolor{MyLigthBlack}{rgb}{0.2,0.25,0.3} 

\hypersetup{%
bookmarks=true
  pdftitle = {Title Here},
  pdfsubject = {},
  pdfkeywords = {Keywords},
  pdfauthor = {Alessandro Casini},}
\hypersetup{ 
  colorlinks = {true}, 
  citecolor = {MyBlue},
  linkcolor = {MyRed},
  filecolor= {Green},	
  urlcolor = {MyRed},
  hyperindex = {true},
  linktocpage = {true},
  linkbordercolor ={1 0 0},
 citebordercolor ={0 1 1},
 urlbordercolor ={0 1 1}
}

\makeatother

\providecommand{\algorithmname}{Algorithm}
\providecommand{\assumptionname}{Assumption}
\providecommand{\conditionname}{Condition}
\providecommand{\corollaryname}{Corollary}
\providecommand{\definitionname}{Definition}
\providecommand{\lemmaname}{Lemma}
\providecommand{\propositionname}{Proposition}
\providecommand{\theoremname}{Theorem}

\begin{document}
\pagebreak{}

\setcounter{page}{0}
\title{\textbf{Continuous Record Laplace-based Inference about the Break
Date in Structural Change Models}\thanks{We are grateful to the Guest Editor, Timothy Vogelsang, for useful
detailed comments. We also thank Zhongjun Qu and Viktor Todorov for
helpful discussions. }\textbf{ }}
\maketitle
\begin{abstract}
Building upon the continuous record asymptotic framework recently
introduced by \citet{casini/perron_CR_Single_Break} for inference
in structural change models, we propose a Laplace-based (Quasi-Bayes)
procedure for the construction of the estimate and confidence set
for the date of a structural change. It is defined by an integration
rather than an optimization-based method. A transformation of the
least-squares criterion function is evaluated in order to derive a
proper distribution, referred to as the Quasi-posterior. For a given
choice of a loss function, the Laplace-type estimator is the minimizer
of the expected risk with the expectation taken under the Quasi-posterior.
Besides providing an alternative estimate that is more precise\textemdash lower
mean absolute error (MAE) and lower root-mean squared error (RMSE)\textemdash than
the usual least-squares one, the Quasi-posterior distribution can
be used to construct asymptotically valid inference using the concept
of Highest Density Region. The resulting Laplace-based inferential
procedure is shown to have lower MAE and RMSE, and the confidence
sets strike a better balance between empirical coverage rates and
average lengths of the confidence sets relative to traditional long-span
methods, whether the break size is small or large.
\end{abstract}

\onehalfspacing
\thispagestyle{empty}
\allowdisplaybreaks

\vfill{}

\pagebreak{}

\section{Introduction}

In recent work \citet{casini/perron_CR_Single_Break}, henceforth
CP, developed a continuous record asymptotic framework for inference
about the break date in a linear time series regression model with
a single shift in some regression parameters while \citet{casini/perron_SC_BP_Lap}
introduced the Generalized Laplace (GL) estimation and inference as
an alternative to least-squares in structural change models. In this
paper we use the GL method under the continuous record asymptotic
framework of CP to tackle some of the problems in this literature.
To illustrate the issues involved, consider the simple model $Y_{kh}=(\delta_{1}^{0}+\delta^{0}\mathbf{1}_{\left\{ k>T_{b}^{0}\right\} })Z_{kh}+e_{kh}.$
We have access to $T+1$ observations $\left(k=0,\ldots,\,T\right)$
at equidistant time intervals $h$ over a fixed time span $N=Th.$
The parameter $\delta^{0}$ is referred to as the magnitude of the
break. The statistical problem is to 1) estimate the break date $T_{b}^{0}$
with highest precision possible and, 2) construct confidence sets
for this quantity that possess correct empirical coverage rates and
short average lengths in finite samples. So far with respect to 1)
little work has been done to consider alternatives to  OLS in the
context of the linear model. With respect to 2) several methods have
been investigated, some of which work well with small breaks, while
others work well with large ones, none of which deliver adequate coverage
rates and ``decent'' lengths of the confidence sets over all break
sizes. We refer to \citet{casini/perron:OUP-Breaks}, \citet{chang/perron:18}
and the discussion later for an investigation on the drawbacks of
popular existing methods, such as those proposed in \citet{bai:97RES}
and \citet{elliott/mueller:07}. In contrast, CP developed an alternative
asymptotic framework based on a continuous record of observations
over a fixed time horizon $\left[0,\,N\right]$. This involves letting
the sample size $T$ grow to infinity by requiring the sampling $h\downarrow0$
at the same rate so that $N=Th$ remains fixed. The limiting distribution
depends on structural parameters but feasible inference can be developed.

\begin{center}
\begin{figure}[h]
\includegraphics[width=16cm,height=7cm]{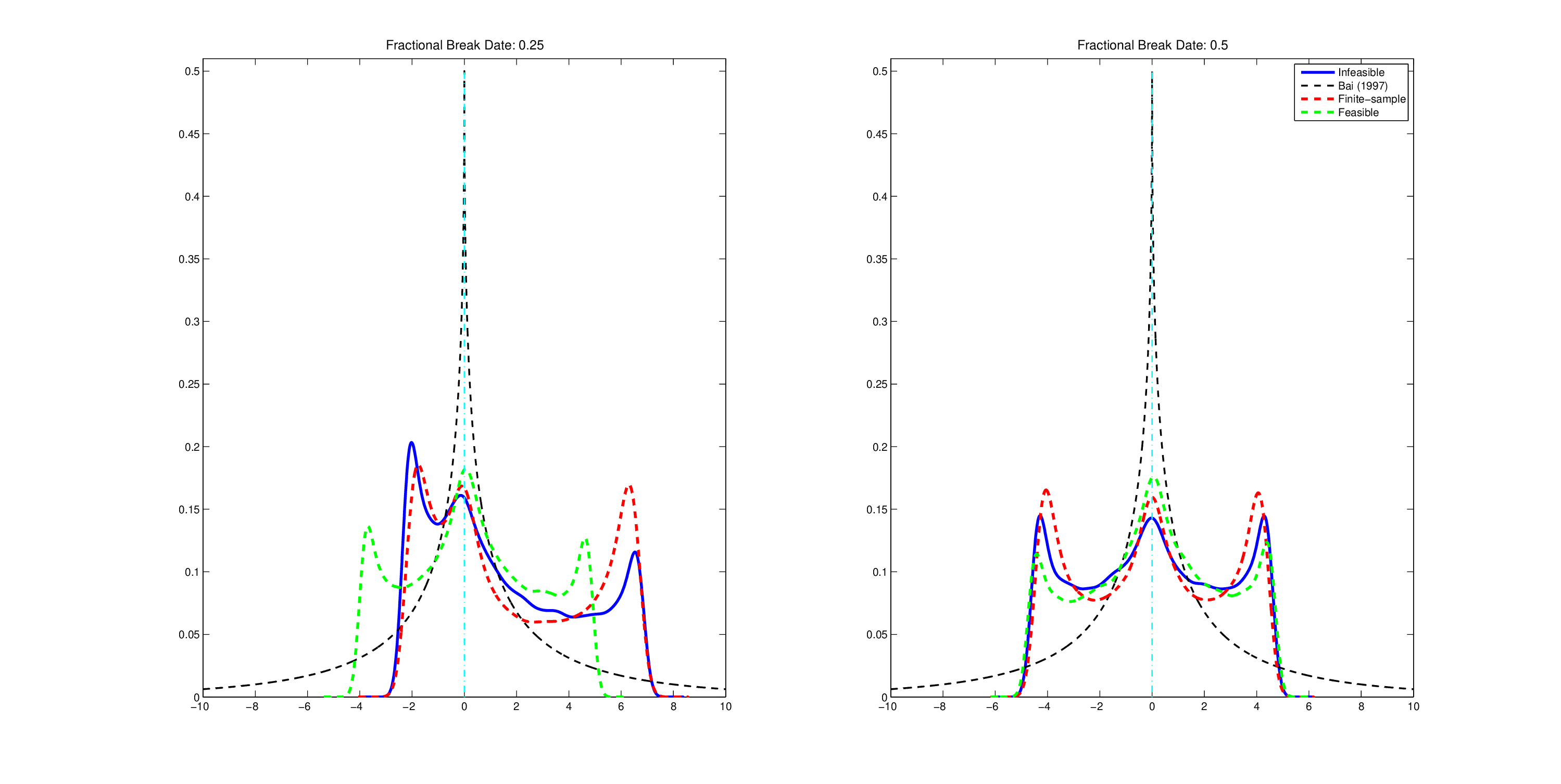}

{\footnotesize{}\caption{{\small{}\label{Fig1}}{\footnotesize{}The probability density of
the least-squares estimator $\widehat{T}_{b}-T_{b}^{0}$ for the model
$y_{t}=\mu^{0}+Z_{t}\delta_{1}^{0}+Z_{t}\delta^{0}\mathbf{1}_{\left\{ t>\left\lfloor T\lambda_{0}\right\rfloor \right\} }+e_{t},\,Z_{t}=0.3Z_{t-1}+u_{t}-0.1u_{t-1},\,u_{t}\sim\textrm{i.i.d.}\mathscr{N}\left(0,\,1\right),\,e_{t}\sim\textrm{i.i.d.}\mathscr{N}\left(0,\,1\right),$
$\left\{ u_{t}\right\} $ independent from $\left\{ e_{t}\right\} ,$
$T=100$ with break magnitude $\delta^{0}=0.3$ and true break point
$\lambda_{0}=0.25$ and $0.5$ (the left and right panel, respectively).
The blue solid (resp., green broken) line is the density of the infeasible
(reps., feasible) continuous record asymptotic distribution of CP,
the black broken line is the density of the asymptotic distribution
from Bai (1997) and the red broken line break is the density of the
finite-sample distribution.}}
}{\footnotesize\par}
\end{figure}
\end{center}

\begin{center}
\begin{figure}[h]
\includegraphics[width=16cm,height=7cm]{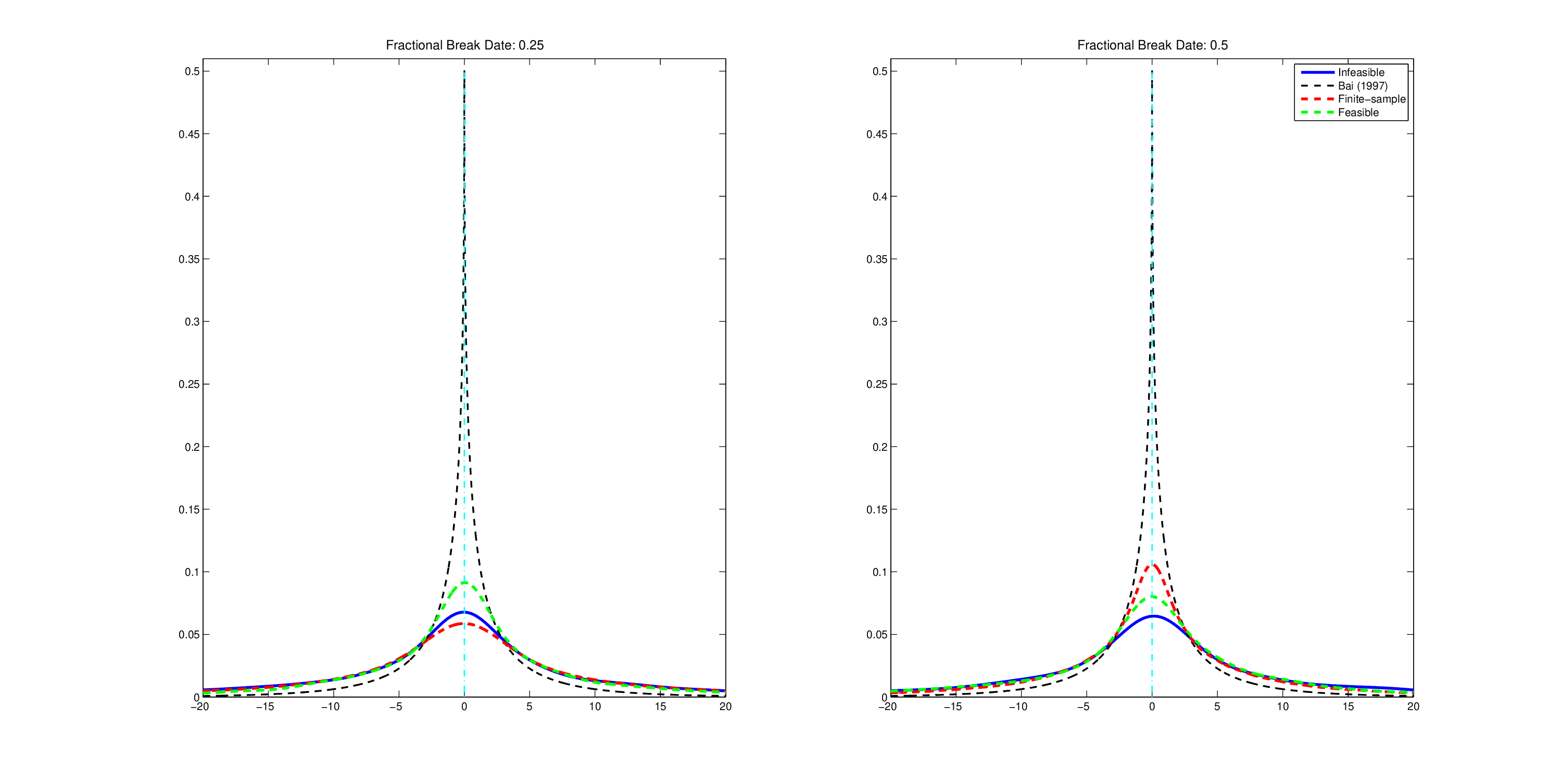}

{\footnotesize{}\caption{{\footnotesize{}\label{Fig2}The comments in Figure \ref{Fig1} apply
but with a break magnitude $\delta^{0}=1.5$.}}
}{\footnotesize\par}
\end{figure}
\end{center}

We consider the asymptotics with $h\downarrow0$ as an alternative
asymptotic experiment that can deliver better approximations for discrete-time
applications and not necessarily as a framework that is useful only
for high-frequency data. Figures \ref{Fig1}-\ref{Fig2} present
plots of the density of the distribution of the least-squares estimate
of the break date for the simple model discussed above with a parameter
shift associated to a regressor $Z_{kh}$ specified to follow an ARMA(1,1)
process and i.i.d. Gaussian disturbances $e_{kh}$. The distributions
presented are: the exact finite-sample distribution, \citeauthor{bai:97RES}\textquoteright s
(1997) classical large-$N$ limit distribution, CP's continuous record
limit distribution and its feasible version (which uses plug-in 
estimates for the true parameter values). The continuous record limiting
distributions provide an impressively accurate approximation to the
finite-sample distribution for both small (cf. Figure \ref{Fig1})
and large break sizes (cf. Figure \ref{Fig2}) and does so for different
break locations (at one-fourth and half-sample in, respectively, the
left and right plots). Additionally, it is evident that the classical
shrinkage large-$N$ asymptotic distribution does not offer an accurate
approximation, especially when the break magnitude is small (cf. Figure
\ref{Fig1}).

\citeauthor{bai:97RES}'s (1997) method for the construction of the
confidence intervals for $T_{b}^{0}$ relies on standard asymptotic
arguments with the span increasing such that each regime increases
proportionately. This coupled with the standard mixing assumption
implies that the limit distribution depends only on a neighborhood
around the true value. This is in stark contrast to the finite-sample
distribution which, especially for small breaks, is influenced by
the location of the break and the properties of the processes over
the whole sample. Hence, it should not be surprising that the confidence
intervals from \citeauthor{bai:97RES}'s (1997) method display empirical
coverage rates often below the nominal level when the break is not
large. CP documented the highly non-standard features of the density
and consequently proposed an alternative inference procedure based
on the concept of Highest Density Region (HDR). Their method was shown
to offer the best balance between empirical coverage rates and average
lengths of the confidence sets among existing methods.

We develop additional inference procedures for the break date in a
linear regression model based on the continuous record asymptotic
framework and using the Generalized Laplace estimation method. The
aim is to provide both a more precise estimate and methods to construct
confidence sets with good finite-sample properties. The idea behind
the Laplace (or Quasi-Bayes) estimator goes back to \citet{laplace:74}.
For a recent application, see \citet{chernozhukov/hong:03} for regular
problems oriented toward microeconometric applications {[}see also
\citet{forneron/ng:17} for a review and comparisons{]}. Laplace-type
estimators rely on statistical criterion functions and integration-based
estimation rather than optimization. They  are constructed as integral
transformations of extremum criterion functions and can be computed
using Markov Chain Monte Carlo methods (MCMC). The integral transformation
of the criterion function is meant to provide an approximation to
the likelihood approach. Hence, they are often referred to as Quasi-Bayesian
estimators. Since no parametric likelihood is available, such a transformation
turns the objective function into a proper density\textemdash referred
to as the Quasi-posterior\textemdash over the parameter of interest.
Our goal is to consider Laplace estimators and HDR methods to construct
the confidence sets, as alternatives to the least-squares estimator
coupled with classical inference procedure, which have better finite
sample properties, namely reduced Mean Absolute Error (MAE) and Root-Mean
Squared Error (RMSE), and confidence sets with accurate coverage rates
and short lengths. Of particular importance, we aim to achieve this
goal whatever the size of the break, a notoriously difficult problem
in the structural change literature.

The use of Laplace-type estimators in \citet{chernozhukov/hong:03}
was mainly motivated by a need to circumvent the curse of dimensionality
inherent in the computation of some classical extremum estimators.
Furthermore, they focused on regular problems for which a quadratic
expansion of the criterion function is available. Our implementation
stems from different concerns. First, our motivation does not arise
from a practical need of reducing the curse of dimensionality inherent
to computation. Indeed, the least-squares estimates are simple to
compute even in models with multiple changes {[}cf. \citet{hawkins:76}
and \citet{bai/perron:03} for an efficient algorithm based on the
principle of dynamic programming{]}. Using the Laplace-type estimator,
we aim to have better estimates than those based on the least-squares
principle; e.g., reduced MAE and RMSE. This should not come at the
expense of less adequate confidence sets. Hence, we use the Laplace-type
estimator in conjunction with the continuous record limiting distribution
and the HDR method to construct confidence sets. A second crucial
distinction  is that estimating the date of a structural change is
a non-standard statistical problem. The limiting distribution of the
break point estimator is related to the location of the maximum of
a two-sided Gaussian process with drift and thus quite different from
the asymptotic distribution of regular estimators.  More importantly,
the objective function does not admit a quadratic expansion, which
adds technical complexities. Although the definition of the Laplace
estimator in our context is similar to \citet{chernozhukov/hong:03},
its asymptotic properties are different in that the contribution of
information provided by the prior function does not vanish asymptotically
due to the non-regularities of the structural change problem. Hence,
the estimator does not share the same Quasi-Bayesian characterization
used by \citet{chernozhukov/hong:03} in regular problems. In addition,
our estimator also differs from the Expectation-Maximization (EM)
algorithm {[}see \citet{dempster/laird/rubin:77}{]} because we compute
the expected risk for a given loss function and not just the expected
value of some log-likelihood function. However, what our approach
shares with the EM algorithm is that when the ``Quasi-prior'' defined
below is used, then our estimator also relies on a two-step procedure.

The Laplace-type estimation is explored as follows. We begin with
the least-squares estimation of the break date and present the feasible
limit distribution developed by CP under a continuous record framework.
We treat the corresponding density function as our \textquotedblleft prior\textquotedblright{}
and call it the \textquotedblleft Quasi-prior\textquotedblright .\footnote{This distribution should not be viewed as a prior in Bayesian sense.
Indeed, it should be simply interpreted as a weight function. Furthermore,
it should be noted that this step implicitly relies on an optimization
procedure which may not be present under the setting of \citet{chernozhukov/hong:03}.} We compute a transformation of the least-squares criterion function
and combine it with the Quasi-prior in order to derive a proper distribution
which we refer to as the \textquotedblleft Quasi-posterior\textquotedblright .
We then use simple computational methods based on integration to define
the so-called \textquotedblleft Generalized Laplace\textquotedblright{}
(GL) estimator. Asymptotically valid inferences are constructed using
HDR-type methods.  The main idea behind the usefulness of our method
can be explained as follows. First, note that the least-squares objective
function is quite flat with respect to the candidates break dates
$T_{b}$. However, the distribution of the least-squares estimate
is quite informative in that it assigns sharply different density
mass across $T_{b}$ (c.f. the multi-modal feature). Hence, working
with the objective function weighted by a Quasi-prior set equal to
the density of the least-square estimate yields estimates with better
properties as shown in the simulations. Using a least-absolute deviation
loss function, the Laplace estimate with our chosen quasi-prior is
shown to have substantially lower MAE and RMSE in finite-samples compared
to that obtained using least-squares. As in other methods, our procedure
relies on a trimming parameter $0<\epsilon<1$ which prevents the
estimator to locate the break date in the first and last $100\epsilon\%$
of the sample. To achieve good results it is necessary that the trimming
parameter should not be chosen too high because otherwise the estimate
might tend to overestimate (resp. underestimate) the break date if
it is in the first (resp. second) half of the sample.  With regards
to the GL estimator, the smaller the trimming the larger is the information
used by the procedure. Since its construction relies on the overall
behavior of the criterion function  we find that $\epsilon=0.05$
performs well for different locations of the break date. Our proposed
inference results in confidence sets for the break date with more
accurate coverage rates and often accompanied by a non-trivial reduction
in the width of the confidence set relative to popular methods. 

Recently \citet{baek:19} proposed to modify the least-squares objective
function by applying some weights so as to reduce the multimodality
effect. The latter estimator displays lower RMSE but has often higher
bias relative to the original LS estimator. The latter may be due
to the fact that the weighting  is not data-dependent while our approach
fully exploits the information from the criterion function since the
weighting or quasi-prior is data-dependent.

This paper relates to two recent papers, namely CP\nocite{casini/perron_CR_Single_Break}
(2020a, 2020b)\nocite{casini/perron_SC_BP_Lap}.  CP\nocite{casini/perron_CR_Single_Break}
developed the continuous record asymptotics that we use in this paper
and proposed confidence sets for the break date based on the continuous
record asymptotic distribution. Hence, it contains the essential ingredients
for our proposed GL method, while at the same time offering alternative
inference procedures. \citet{casini/perron_SC_BP_Lap} analyzed the
GL method under classical asymptotics and focused on the theoretical
relationship between the asymptotic distribution of frequentist and
Bayesian estimators of the break point. They showed that depending
on some input parameter, the GL estimator exhibits a dual limit distribution\textemdash the
shrinkage asymptotic distribution of \citet{bai/perron:98}, or a
Bayes-type asymptotic distribution {[}cf. \citet{ibragimov/has:81}{]}.
Moreover, the results in the former extend to models with multiple
breaks and models with trending regressors which are not covered in
this paper. Finally, some of the results in this paper allow for long-memory
which is ruled out by \citet{casini/perron_SC_BP_Lap}. See also \citet{phillips:87a}
and \citet{perron:91} for early applications of continuous record
asymptotics in the context of unit roots.

Recent works have used continuous-time asymptotics for structural
change and nonstationarity models more generally. Besides the already
discussed approach of CP, \citet{jiang/wang/yu:16} studied the finite-sample
bias of a break point estimator for a univariate diffusion with constant
volatility and a change-point in the drift while \citet{jiang/wang/yu:17}
focused on the break point estimator for the Ornstein-Uhlenbeck process.
Their approach is different from CP in that their results are less
general and feasible inference for the break point is not discussed.
Finally, \citet{chambers/taylor:19} considered both deterministic
one-time and continuous stochastic parameter change in a continuous-time
autoregressive model while \citet{casini_CR_Test_Inst_Forecast} introduced
continuous-time asymptotics to test for forecast failure.

The paper is organized as follows. Section \ref{Section: Statistical Setting LapCR}
introduces the setup. Section \ref{Sec Continuous-Record-Asymptotic Results}
summarizes results from CP needed for subsequent analyses. In Section
\ref{Section Asymptotic Results LapCR}, we develop the large-sample
properties of the GL estimators. We verify their accuracy in Section
\ref{sec:Small-Sample-Properties-of}. Section \ref{Section Inference-Methods-based On Laplace}
describes the inference methods proposed, which are evaluated via
simulations Section \ref{Section Monte Carlo Study}. Section \ref{Section Conclusive-Remarks}
offers brief concluding remarks. All technical derivations are contained
in a supplement {[}\citet{casini/perron_Lap_CR_single_Inf_Supp}{]}.

\section{The Statistical Setting\label{Section: Statistical Setting LapCR}}

Section \ref{Section, The Model} introduces the model and some assumptions.
We also provide a discussion of our assumptions and the relationship
of our framework with that of the long-span (shrinkage) asymptotics
considered by \citet{bai:97RES}, \citet{bai/perron:98} and \citet{perron/qu:06}.
The least-squares estimator is defined in Section \ref{Subsection LS Estimator}.
The following notations are used in the sequel. All limits are taken
as $T\rightarrow\infty$ with the span $N$ kept fixed, where $N=Th$
with $h$, the sampling interval so that $h\downarrow0$ at the same
rate as $T\rightarrow\infty$. $\mathbb{R}$ denotes the set of real
numbers. For two vectors $a$ and $b$, we write $a\leq b$ if the
inequality holds component-wise. We denote the transpose of a matrix
$A$ by $A'$. The $\left(i,\,j\right)$ elements of $A$ are denoted
by $A^{\left(i,j\right)}$. We use $\left\Vert \cdot\right\Vert $
to denote the Euclidean norm of a linear space i.e., $\left\Vert x\right\Vert =\left(\sum_{i=1}^{p}x_{i}^{2}\right)^{1/2}$
for $x\in\mathbb{R}^{p}.$ For a matrix $A$ we use the vector-induced
norm, i.e., $\left\Vert A\right\Vert =\sup_{x\neq0}\left\Vert Ax\right\Vert /\left\Vert x\right\Vert .$
For a sequence of matrices $\left\{ A_{T}\right\} ,$ we write $A_{T}=o_{\mathbb{P}}\left(1\right)$
if each of its elements is $o_{\mathbb{P}}\left(1\right)$ and likewise
for $O_{\mathbb{P}}\left(1\right).$ The symbol $\left\lfloor \cdot\right\rfloor $
denotes the largest smaller integer function while $\otimes$ is used
for the product of $\sigma$-fields. A sequence $\left\{ u_{kh}\right\} _{k=1}^{T}$
is $\textrm{i.i.d.}$ (resp., $\textrm{i.n.d.}$) if the $u_{kh}$
are independent and identically (resp., non-identically) distributed.
We use $\overset{\mathbb{P}}{\rightarrow}$, and $\Rightarrow$ to
denote, respectively, convergence in probability (under some measure
$\mathbb{P}$), and weak convergence, while $\overset{\mathcal{L}-\textrm{s}}{\Rightarrow}$
is used to denote stable convergence in law. We write $\mathscr{MN}\left(a,\,b\right)$
for the mixed normal distribution with parameters $\left(a,\,b\right)$.
The space $\mathbb{M}_{p}^{\mathrm{c\grave{a}dl\grave{a}g}}$ collects
all $p\times p$ positive definite real-valued matrices whose elements
are \textit{càdlàg}. We use the superscript $+$ in $\mathbb{M}_{p}^{\mathrm{}}$
if these matrices are required to be positive definite. $\boldsymbol{FV}^{\mathrm{c}}$
denotes the class of continuous adapted finite variation processes
and $\boldsymbol{M}_{\mathrm{loc}}^{\mathrm{c}}$ denotes the class
of continuous local martingale processes with finite positive definite
conditional variance. For semimartingales $\left\{ S_{t}\right\} _{t\geq0}$
and $\left\{ V_{t}\right\} _{t\geq0}$, we denote their covariation
process by $\left[S,\,V\right]_{t}$ and its predictable counterpart
by $\left\langle S,\,V\right\rangle _{t}$. We anticipate that in
our setting the latter two processes will be equivalent. The symbol
``$\triangleq$'' refers to definitional equivalence.

\subsection{The Model\label{Section, The Model}}

A standard discrete-time partial structural change model with a single
break is given by:
\begin{flalign}
 &  & Y_{t} & =D_{t}'\varrho^{0}+Z_{t}'\delta_{1}^{0}+u_{t},\qquad\qquad\qquad\qquad\qquad\qquad\left(t=0,\ldots,\,T_{b}^{0}\right)\qquad\qquad\label{Original SC Model}\\
 &  & Y_{t} & =D_{t}'\varrho^{0}+Z_{t}'\delta_{2}^{0}+u_{t},\qquad\qquad\qquad\qquad\qquad\qquad\left(t=T_{b}^{0}+1,\ldots,\,T\right),\qquad\qquad\nonumber 
\end{flalign}
where $Y_{t}$ is  the dependent variable, $D_{t}$ and $Z_{t}$
are, respectively, $p\times1$ and $q\times1$ vectors of regressors
and $e_{t}$ is an unobservable disturbance. The statistical problem
is to estimate the unknown  parameters $\varrho^{0},\,\delta_{1}^{0}$,
$\delta_{2}^{0}$ and the break date $T_{b}^{0}$. A structural change
occurs at $T_{b}^{0}$ because by assumption $\delta_{1}^{0}\neq\delta_{2}^{0}$.
The break magnitude is $\delta^{0}=\delta_{2}^{0}-\delta_{1}^{0}$.
The inference problem is to construct estimates and confidence sets
for the break date $T_{b}^{0}$ when $T+1$ observations on $\left(Y_{t},\,D_{t},\,Z_{t}\right)$
are available. 

We introduce the high-frequency setting which serves as a probabilistic
background on which our asymptotic arguments are developed. There
is a filtered probability space $\left(\Omega,\,\mathscr{F},\,\left(\mathscr{F}_{t}\right)_{t\geq0},\,\mathbb{P}\right)$
on which the continuous time process $X\triangleq\left(Y,\,D',\,Z'\right)'$
is defined, where $Y\triangleq\left\{ Y_{t}\right\} _{t\geq0},$ $D\triangleq\left\{ D_{t}\right\} _{t\geq0},$
$Z\triangleq\left\{ Z_{t}\right\} _{t\geq0}$ are assumed to be It\^{o}
semimartingales. We observe $T+1$ realizations of $Y_{t},\,D_{t}$
and $Z_{t}$ at equidistant discrete times $t=0,\,h,\,2h,\ldots,\,N$,
where $N\triangleq Th$ is the length of the fixed time span $\left[0,\,N\right]$.
Note that in general $N$ is not identified and could be normalized
to one. However, we keep a generic $N$ throughout to allow a better
intuitive understanding of the results. Under the continuous record
asymptotic scheme, we let the sample size $T$ grow by shrinking the
sampling interval $h$ to zero so that the span $N$ remains fixed. 

For each $h,$ $D_{kh}\in\mathbb{R}^{p}$ and $Z_{kh}\in\mathbb{R}^{q}$
are random vector step functions changing at times $0,\,h,\,2h,\ldots,\,Th$.
The continuous-time unobservable error sequence is the \textit{càdlàg}
adapted process $\left\{ e_{t}^{*}\right\} $. The regressors $D_{kh}$
and $Z_{kh}$ include locally-integrable semimartingales (extended
to allow for predictable processes in Section \ref{subsection The-Extended-Model}).
We assume that the discretized processes $D_{kh}$ and $Z_{kh}$
are adapted to the increasing and right-continuous filtration $\left\{ \mathscr{F}_{t}\right\} $.
By the Doob-Meyer decomposition we can write,\footnote{For any process $X$ we denote the ``increments'' of $X$ by $\Delta_{h}X_{k}=X_{kh}-X_{\left(k-1\right)h}$.
We sometimes use $\Delta t$ in place of $h$ in order to conveniently
make our sums look like integrals. } for $k=1,\ldots,T$: $\Delta_{h}D_{k}\triangleq\mu_{D,k}h+\Delta_{h}M_{D,k}$
and $\Delta_{h}Z_{k}\triangleq\mu_{Z,k}h+\Delta_{h}M_{Z,k}$, where
the drifts $\mu_{D,t}\in\mathbb{R}^{p},\,\mu_{Z,t}\in\mathbb{R}^{q}$
are $\mathscr{F}_{t-h}$-measurable, and $M_{D,k}\in\mathbb{R}^{p},\,M_{Z,k}\in\mathbb{R}^{q}$
are continuous local martingales with $\mathbb{P}$-a.s. finite conditional
covariance matrices $\mathbb{E}\left(\Delta_{h}M_{D,t}\Delta_{h}M_{D,t}'|\,\mathscr{F}_{t-h}\right)=\Sigma_{D,t-h}\Delta t$
and $\mathbb{E}\left(\Delta_{h}M_{Z,t}\Delta_{h}M'_{Z,t}|\,\mathscr{F}_{t-h}\right)=\Sigma_{Z,t-h}\Delta t$.
Exact assumptions will be given below. To the continuous-time process
$\left\{ e_{t}^{*}\right\} $ corresponds a disturbance sequence $\left\{ \Delta_{h}e_{t}^{*}\right\} $
such that $\left\{ \Delta_{h}e_{t}^{*},\,\mathscr{F}_{t}\right\} $
is a continuous local martingale difference sequence with finite
conditional variance given by $\mathbb{E}\left[\left(\Delta_{h}e_{t}^{*}\right)^{2}|\,\mathscr{F}_{t-h}\right]=\sigma_{e,t-h}^{2}\Delta t$
$\mathbb{P}$-a.s. Under this formulation, the discretized model
is 
\begin{align}
\Delta_{h}Y_{k} & \triangleq\begin{cases}
\left(\Delta_{h}D_{k}\right)'\varrho^{0}+\left(\Delta_{h}Z_{k}\right)'\delta_{1}^{0}+\Delta_{h}e_{k}^{*},\qquad & \left(k=1,\ldots,\left\lfloor T\lambda_{0}\right\rfloor \right)\\
\left(\Delta_{h}D_{k}\right)'\varrho^{0}+\left(\Delta_{h}Z_{k}\right)'\delta_{2}^{0}+\Delta_{h}e_{k}^{*},\qquad & \left(k=\left\lfloor T\lambda_{0}\right\rfloor +1,\ldots,\,T\right)
\end{cases}\label{Eq. Model 1, CT}
\end{align}
 where $\lambda_{0}\in\left(0,\,1\right),\,\varrho^{0}\in\mathbb{R}^{p},\,\delta_{1}^{0},\,\delta_{2}^{0}\in\mathbb{R}^{q},$
and $\delta^{0}\triangleq\delta_{2}^{0}-\delta_{1}^{0}$. It holds
that $\delta^{0}\neq0$ so that a structural change in the parameter
associated with $\Delta_{h}Z_{k}$ has occurred at time $\left\lfloor N\lambda_{0}\right\rfloor $.
Under our setting, we need to distinguish between the actual break
date $N_{b}^{0}=N\lambda_{0}$ (on a calendar time) and the index
of the discrete-time observation associated with the break point:
$T_{b}^{0}\triangleq\left\lfloor T\lambda_{0}\right\rfloor =\left\lfloor N_{b}^{0}/h\right\rfloor $.
Under the classical long-span setting this is not necessary since
$h=1$ or $N=T$. 

The specification implicitly assumes the following continuous-time
data-generating process,
\begin{align}
D_{t}=D_{0}+\int_{0}^{t}\mu_{D,s}ds+\int_{0}^{t}\sigma_{D,s}dW_{D,s}, & \qquad Z_{t}=Z_{0}+\int_{0}^{t}\mu_{Z,s}ds+\int_{0}^{t}\sigma_{Z,s}dW_{Z,s},\label{Model Regressors Integral Form}
\end{align}
where $\mu_{D,t}$ (resp., $\mu_{Z,t}$) is the infinitesimal conditional
mean of $D_{t}$ (resp., $Z_{t}$) and takes value in $\mathbb{\mathbb{R}}^{p}$
(resp., $\mathbb{R}^{q}$); $\sigma_{D,t}$ and $\sigma_{Z,t}$ are
the instantaneous covariance processes taking values in $\mathbb{M}_{p}^{\textrm{càdlàg}}$
and $\mathbb{M}_{q}^{\textrm{càdlàg}}$, respectively; $W_{D}$ (resp.,
$W_{Z}$) is a $p$ (resp., $q$)-dimensional standard Wiener process;
and $D_{0}$ and $Z_{0}$ are $\mathscr{F}_{0}$-measurable random
vectors. The process $e_{t}^{*}$ is a continuous local martingale
orthogonal (in martingale sense) to both $D_{t}$ and $Z_{t}$, i.e.,
$\left\langle e,\,D\right\rangle _{t}=\left\langle e,\,Z\right\rangle _{t}=0$
for all $t$. We consider only processes with continuous sample paths,
as stated in the following assumption. 
\begin{assumption}
\label{Assumption Continuous Sample Path LapCR}$D,\,Z,\,e$ and $\Sigma^{0}\triangleq\left\{ \Sigma_{\cdot,t},\,\sigma_{e,t}\right\} _{t\geq0}$
are continuous Itô semimartingales.
\end{assumption}
Assumption \ref{Assumption Continuous Sample Path LapCR} rules out
processes with discontinuous sample paths from our analysis. Hence,
our results are not expected to provide good approximations for applications
involving high-frequency data for which jumps are likely to be important;
e.g., \citet{andersen/fusari/todorov:16}, \citet{bandi/reno:16}
and \citet{hounyo/liu/varneskov:20} for financial-oriented applications
and \citet{li/xiu:16} for a GMM setup; for a textbook account, see
\citet{sahalia/jacod:14}. Another important difference from the high-frequency
statistics centers on the \textquotedblleft mean effect\textquotedblright{}
which is captured here by the drift process. This poses several challenges
because the drift is not identified under a continuous record asymptotics.
CP dealt with this issue by introducing a so-called small-dispersion
model, for which the parameters affecting the limit distribution can
be consistently estimated and all theoretical results are shown to
apply to this case as well (c.f. Section \ref{subsection The-Extended-Model}
below).
\begin{assumption}
\label{Assumption 1, CT}The model \eqref{Eq. Model 1, CT}-\eqref{Model Regressors Integral Form}
satisfies the following: (i) the càdlàg processes $\sigma_{D,t}$
and $\sigma_{Z,t}$ are locally bounded; (ii) $\mu_{D,t}$ and $\mu_{Z,t}$
are locally bounded. $\int_{0}^{t}\mu_{D,s}ds$ and \textup{$\int_{0}^{t}\mu_{Z,s}ds$
}belong to $\boldsymbol{FV}^{\mathrm{c}}$; (iii) $\int_{0}^{t}\sigma_{D,s}dW_{D,s},\,\int_{0}^{t}\sigma_{Z,s}dW_{Z,s}\in\boldsymbol{M}_{\mathrm{loc}}^{\mathrm{c}}$.
Further, they possess $\mathbb{P}$-a.s. finite positive definite
conditional variances defined by $\Sigma_{D,t}=\sigma_{D,t}\sigma'_{D,t}$
and $\Sigma_{Z,t}=\sigma_{Z,t}\sigma'_{Z,t},$ which for all $t<\infty$
satisfy, for $j=1,\ldots,\,p$: $\int_{0}^{t}\Sigma_{D,s}^{\left(j,j\right)}ds<\infty$
and $\int_{0}^{t}\Sigma_{Z,s}^{\left(j,j\right)}ds<\infty$ where
$\Sigma_{D,t}^{\left(j,r\right)}$ denotes the $\left(j,r\right)$-th
element of the process $\Sigma_{D,t}$. Furthermore, for every $j=1,\ldots,\,p,$
$r=1,\ldots,\,q$, and $k=1,\ldots,\,T$, $h^{-1}\int_{\left(k-1\right)h}^{kh}\Sigma_{D,s}^{\left(j,j\right)}ds$
and $h^{-1}\int_{\left(k-1\right)h}^{kh}\Sigma_{Z,s}^{\left(r,r\right)}ds$
are finite and bounded away from zero, uniformly in $k$ and $h$;
(iv) $e^{*}\triangleq\left\{ e_{t}^{*}\right\} _{t\geq0}$ is a continuous
local martingale satisfying $e_{t}^{*}\triangleq\int_{0}^{t}\sigma_{e,s}dW_{e,s}$
with $0<\sigma_{e,t}^{2}<\infty$ for all $t\geq0$, where $W_{e}$
is a Wiener process. Also, $\left\langle e,\,D\right\rangle _{t}=\left\langle e,\,Z\right\rangle _{t}=0$,
$t\geq0.$
\end{assumption}
Parts (i)-(iii) contains regularity conditions imposed in the high-frequency
financial statistics literature {[}cf. \citet{barndorff/shephard:04},
\citet{li/todorov/tauchen:17} and \citet{li/xiu:16}{]}. Part (iv)
specifies the disturbance process to possess continuous sample paths
and to be contemporaneously orthogonal to the regressors. Assumptions
\ref{Assumption Continuous Sample Path LapCR}-\ref{Assumption 1, CT}
imply that the processes in our model are diffusion-type processes
if one further assumes that the volatilities are deterministic. We
shall not impose the latter condition. As a consequence, $X=\left(Y,\,D',\,Z'\right)'$
is a member of the continuous stochastic volatility semimartingale
class. We model volatility as a latent factor since it does not pose
any substantial impediment for the development of our theoretical
results and they are more general including, for example, nonstationarity
and long-memory. 
\begin{assumption}
\label{Assumption 3 Break Date} $N_{b}^{0}=\left\lfloor N\lambda_{0}\right\rfloor $,
for some $\lambda_{0}\in\left(0,\,1\right).$ 
\end{assumption}
Assumption \ref{Assumption 3 Break Date} dictates the asymptotic
framework adopted and implies that the break date occurs at the observation-index
$T_{b}^{0}=T\lambda_{0}$, where $T_{b}^{0}=N_{b}^{0}/h$.  Under
our framework it implies that the pre- and post-break segments of
the sample remain fixed. The usual assumption under the classical
large-$N$ asymptotics implies that the time horizons before and after
the break date grow proportionately, which, along with the mixing
assumption implies that only a small neighborhood around the true
break date is relevant asymptotically, thereby ruling out the possibility
 to discern any asymmetric feature of the asymptotic distribution
simply caused by the location of the break date. The continuous record
asymptotic framework preserves information about the data span and
the location of the break, and the mixing and ergodic assumptions
are not needed. CP showed that the theoretical results derived for
conducting inference about the break date in model \eqref{Eq. Model 1, CT}
are applicable to classical structural change models for which a long-span
setting is usually adopted. It is convenient to use the following
re-parametrization. Let $y_{kh}=\Delta_{h}Y_{kh},$ $x_{kh}=\left(\Delta_{h}D'_{k},\,\Delta_{h}Z'_{k}\right)'$,
$z_{kh}=\Delta_{h}Z_{k}$, $e_{kh}=\Delta_{h}e_{k}^{*}$, and $\beta^{0}=((\varrho^{0})',\,(\delta_{1}^{0})')'$.
Then, \eqref{Eq. Model 1, CT} is
\begin{align}
y_{kh} & =x_{kh}'\beta^{0}+e_{kh},\qquad & \left(k=1,\ldots,\,T_{b}^{0}\right)\label{Model (4), scalar format, CT}\\
y_{kh} & =x_{kh}'\beta^{0}+z_{kh}'\delta^{0}+e_{kh},\qquad & \,\left(k=T_{b}^{0}+1,\ldots,\,T\right).\nonumber 
\end{align}
We further define the full column rank $\left(p+q\right)\times q$
known matrix $H$ such that $z_{kh}=H'x_{kh}$. We can then state
model \eqref{Model (4), scalar format, CT} in matrix format, which
will be used for the derivations. Let $Y=\left(y_{h},\,\ldots,\,y_{Th}\right)',\,X=\left(x_{h},\,\ldots,\,x_{Th}\right)'$,
$e=\left(e_{h},\,\ldots,\,e_{Th}\right)',$ $X_{1}=\left(x_{h},\,\ldots,\,x_{T_{b}h},\,0,\,\ldots,\,0\right)'$,
$X_{2}=\left(0,\,\ldots,\,0,\,x_{\left(T_{b}+1\right)h},\ldots,\,x_{Th}\right)'$
and $X_{0}=(0,\,\ldots,\,0,\,x_{\left(T_{b}^{0}+1\right)h},\ldots,\,x_{Th})'$.
Also, $Z_{1}=X_{1}H,\,Z_{2}=X_{2}H$ and $Z_{0}=XH$. Then \eqref{Model (4), scalar format, CT}
is equivalent to 
\begin{align}
Y & =X\beta^{0}+Z_{0}\delta^{0}+e.\label{Model Matrix format, CT}
\end{align}

\subsection{\label{subsection The-Extended-Model}The Extended Model with Predictable
Processes}

The assumptions on $D_{t}$ and $Z_{t}$ specify that these processes
are continuous Itô semimartingales of the form \eqref{Model Regressors Integral Form}.
This precludes predictable processes, which are often of interest
in applications; e.g., a constant and/or a lagged dependent variable.
Technically, these require a separate treatment since the coefficients
associated with predictable processes are not identified under a fixed-span
asymptotic setting. CP considered the following extended model:
\begin{flalign}
\Delta_{h}Y_{k} & \triangleq\begin{cases}
\mu_{1,h}h+\alpha_{1,h}Y_{\left(k-1\right)h}+\left(\Delta_{h}D_{k}\right)'\varrho^{0}+\left(\Delta_{h}Z_{k}\right)'\delta_{Z,1}^{0}+\Delta_{h}e_{k}^{*},\qquad & \left(k=1,\ldots,\left\lfloor T\lambda_{0}\right\rfloor \right)\\
\mu_{2,h}h+\alpha_{2,h}^ {}Y_{\left(k-1\right)h}+\left(\Delta_{h}D_{k}\right)'\varrho^{0}+\left(\Delta_{h}Z_{k}\right)'\delta_{Z,2}^{0}+\Delta_{h}e_{k}^{*},\qquad & \left(k=\left\lfloor T\lambda_{0}\right\rfloor +1,\ldots,\,T\right)
\end{cases}\label{Model Extended, Sect 1}
\end{flalign}
for some given initial value $Y_{0}$. We specify the parameters associated
with the constant and the lagged dependent variable as increasing
as $h\downarrow0$ in order for some fixed true parameter to remain
in the asymptotics. This is done by specifying: $\mu_{1,h}\triangleq\mu_{1}^{0}h^{-1/2}$,
$\mu_{2,h}\triangleq\mu_{2}^{0}h^{-1/2}$, $\mu_{\delta,h}\triangleq\mu_{2,h}-\mu_{1,h}$,
$\alpha_{1,h}\triangleq\alpha_{1}^{0}h^{-1/2}$, $\alpha_{2,h}\triangleq\alpha_{2}^{0}h^{-1/2}$
and $\alpha_{\delta,h}\triangleq\alpha_{2,h}-\alpha_{1,h}$. Our framework
is then similar to the small-diffusion setting which has been extensively
studied in the statistics literature {[}cf. \citet{ibragimov/has:81},
\citet{laredo:00} and \citet{sorensen/uchida:03}{]}. The results
to be discussed below go through with modifications using the results
in CP. In particular, a two-step procedure is needed to estimate the
parameters as discussed in Section A.2 in CP. The model and results
can be trivially extended to allow more general forms of predictable
processes. 

\subsection{The Least-Squares Estimator of the Break Date\label{Subsection LS Estimator}}

CP considered the break date least-squares (LS) estimator $\widehat{N}_{b}^{\mathrm{LS}}=\widehat{T}_{b}^{\mathrm{LS}}h$
defined as the minimizer of the sum of squares residuals $S_{h}\left(\theta,\,N_{b}\right)$
from \eqref{Model (4), scalar format, CT}, where $\theta\triangleq\left(\beta',\,\delta'\right)'$.
The parameter vector $\theta$ can be concentrated out of the criterion
function to yield an optimization problem over $N_{b}$ only: 
\begin{align*}
\widehat{\theta}_{h}^{\mathrm{LS}}\left(N_{b}\right)=\underset{\theta}{\mathrm{argmin}\,}S_{h}\left(\theta,\,N_{b}\right),\qquad\qquad\widehat{N}_{b}^{\mathrm{LS}}=\underset{hq\leq N_{b}\leq N}{\mathrm{argmin}}\,S_{h}\left(\widehat{\theta}^{\mathrm{LS}}\left(N_{b}\right),\,\left(N_{b}\right)\right) & .
\end{align*}
Using a correspondence between the sum of squared residuals and the
sup-Wald statistic, we have: 
\begin{align}
\underset{hq\leq N_{b}\leq N}{\mathrm{argmin}}\,S_{h}\left(\widehat{\theta}^{\mathrm{LS}}\left(N_{b}\right),\,N_{b}\right) & =\underset{hq\leq N_{b}\leq N}{\mathrm{argmax}}\,\widehat{\delta}^{\mathrm{LS}}\left(N_{b}\right)'\left(Z_{2}'M_{X}Z_{2}\right)\widehat{\delta}^{\mathrm{LS}}\left(N_{b}\right),\label{Eq. Relationship Sup Wald and SSR}
\end{align}
 where $M_{X}\triangleq I-X\left(X'X\right)X'$ and $\widehat{\delta}^{\mathrm{LS}}\left(N_{b}\right)$
is the LS estimator of $\delta^{0}$ obtained by regressing $Y$ on
$X$ and $Z_{2}$. This follows since $Q_{h}\left(\widehat{\theta}^{\mathrm{LS}}\left(N_{b}\right),\,N_{b}\right)\triangleq\widehat{\delta}^{\mathrm{LS}}\left(N_{b}\right)'\left(Z_{2}'M_{X}Z_{2}\right)\widehat{\delta}^{\mathrm{LS}}\left(N_{b}\right)$
is the numerator of a modified sup-Wald statistic. The GL estimator
will depend on the criterion function $Q_{h}\left(\theta\left(N_{b}\right),\,N_{b}\right)$
evaluated at each possible break date. Hence, we need the following
assumptions.
\begin{assumption}
$\theta^{0}=\left(\left(\beta^{0}\right)',\,\left(\delta^{0}\right)'\right)'\in\Theta\subset\mathbb{R}^{\mathrm{dim}\left(\theta\right)}$,
a compact parameter space. 
\end{assumption}

\begin{assumption}
\label{Assumption 4 Eigenvalue}There exists an $l_{0}$ such that
for all $l>l_{0},$ the matrices $\left(lh\right)^{-1}\sum_{k=1}^{l}x_{kh}x'_{kh},$
$\left(lh\right)^{-1}\sum_{k=T-l+1}^{T}x_{kh}x'_{kh},$ $\left(lh\right)^{-1}\sum_{k=T_{b}^{0}-l+1}^{T_{b}^{0}}x_{kh}x'_{kh},$
and $\left(lh\right)^{-1}\sum_{k=T_{b}^{0}+1}^{T_{b}^{0}+l}x_{kh}x'_{kh}$
have minimum eigenvalues bounded away from zero in probability.
\end{assumption}

\begin{assumption}
\label{Assumption 5 Identification}Let $\overline{Q}_{h}\left(\theta^{0},\,N_{b}\right)\triangleq\mathbb{E}\left[Q_{h}\left(\theta^{0},\,N_{b}\right)-Q_{h}\left(\theta^{0},\,N_{b}^{0}\right)\right].$
There exists a $N_{b}^{0}$ such that $\overline{Q}_{h}\left(\theta^{0},\,N_{b}^{0}\right)>\sup_{\left(\theta^{0},\,N_{b}\right)\notin\mathbf{B}}\overline{Q}_{h}\left(\theta^{0},\,N_{b}\right),$
for every open set $\mathbf{B}$ that contains $\left(\theta^{0},\,N_{b}^{0}\right)$. 
\end{assumption}
Assumption \ref{Assumption 4 Eigenvalue} is in the same spirit as
A2 in \citet{bai/perron:98}. It requires that there be enough variation
around the break point so that it can be identified. Multiplying the
increment $x_{kh}$ by the factor $h^{-1/2}$ allows one to normalize
$x_{kh}$ so that the assumption is implied by a weak law of large
numbers. Assumption \ref{Assumption 5 Identification} is a conventional
uniqueness condition.

\section{Asymptotic Results for Least-squares Estimation\label{Sec Continuous-Record-Asymptotic Results}}

In this section, we review continuous record asymptotic results about
the break date LS estimator from CP. They provide intuition on the
forthcoming asymptotic results about the Laplace estimator since both
have a common LS criterion function. The typical asymptotic framework
for structural change problems relies on a shrinking shifts assumption.
Under a continuous record, we also require a small shifts assumption
to analyze the limit distribution of the Laplace estimator. We compare
the asymptotics of CP with that of \citet{bai:97RES} and \citet{elliott/mueller:07}
in Section \ref{Subsection: Comparison CR}.

\subsection{Continuous Record Asymptotics}
\begin{assumption}
\label{Assumption 6 - Small Shifts B}Let $\delta_{h}=h^{1/4}\delta^{0}$
and assume that for all $t\in\left(N_{b}^{0}-\epsilon,\,N_{b}^{0}+\epsilon\right),$
with $\epsilon\downarrow0$ and $T^{1-\kappa}\epsilon\rightarrow B<\infty$,
$0<\kappa<1/2$, $\mathbb{E}\left[\left(\Delta_{h}e_{t}^{*}\right)^{2}|\,\mathscr{F}_{t-h}\right]=\sigma_{h,t-h}^{2}\Delta t$
$\mathbb{P}$-a.s, where $\sigma_{h,t}\triangleq\sigma_{h}\sigma_{e,t}$,
$\sigma_{h}\triangleq h^{-1/4}\overline{\sigma}$ and $\overline{\sigma}\triangleq\int_{0}^{N}\sigma_{e,s}^{2}ds$. 
\end{assumption}
The first part states that the shift parameter converges to zero
at a controlled rate. The second allows for a higher degree of uncertainty
around the change-point by requiring $\left\{ \Delta_{h}e_{t}^{*}\right\} $
to oscillate more as $h\downarrow0$. The latter neither prevents
nor facilitates the identification of the break, i.e., it plays no
role for the global properties of the estimator, namely the consistency
of $\widehat{N}_{b}$ and $\widehat{\theta}$ as well as the asymptotic
distribution of $\widehat{\theta}$, though not that of $\widehat{N}_{b}$.

Under Assumption \ref{Assumption 6 - Small Shifts B}, the rate of
convergence of the LS estimator is $T^{1-\kappa},$ with $0<\kappa<1/2$.
This rate is fast and therefore the volatility of the errors is scaled
up around the change-point so that the objective function behaves
as if it were a standard diffusion process. Note that the $T^{-\left(1-\kappa\right)}$-neighborhood
in which the errors have higher variance arises from the $T^{1-\kappa}$-rate
of convergence of $\widehat{\lambda}_{b}$. Note also that the rate
of convergence $T^{1-\kappa}$ is sufficiently fast to guarantee a
$\sqrt{T}$-consistent estimation of the slope parameters. The results
proved in CP are stated in the following propositions. Let $\Sigma^{*}\triangleq\left\{ \mu_{\cdot,t},\,\Sigma_{\cdot,t},\,\sigma_{e,t}\right\} _{t\geq0}$,
and $Z_{\Delta}\triangleq(0,\ldots,\,0,\,z_{\left(T_{b}+1\right)h},\ldots,\,z_{T_{b}^{0}h},\,0,\ldots,\,0)$
if $T_{b}<T_{b}^{0}$ and $Z_{\Delta}\triangleq(0,\ldots,\,0,\,z_{\left(T_{b}^{0}+1\right)h},\ldots,\,z_{T_{b}h},\,0,\ldots,\,0)$
if $T_{b}>T_{b}^{0}$. Define
\begin{align}
\Delta_{h}\widetilde{e}_{t} & \triangleq\begin{cases}
\Delta_{h}e_{t}^{*}, & t\notin\left(N_{b}^{0}-\epsilon,\,N_{b}^{0}+\epsilon\right)\\
h^{1/4}\Delta_{h}e_{t}^{*}, & t\in\left(N_{b}^{0}-\epsilon,\,N_{b}^{0}+\epsilon\right)
\end{cases}.\label{Eq. eps WN}
\end{align}

\begin{prop}
\label{Prop 3 Asym}Under Assumptions \ref{Assumption 1, CT}-\ref{Assumption 5 Identification}
and \ref{Assumption 6 - Small Shifts B}: (i) $\widehat{N}_{b}\overset{\mathbb{P}}{\rightarrow}N_{b}^{0}$;
(ii) for every $\varepsilon>0$ there exists a $K>0$ such that for
all large $T,$ $\mathbb{P}\left(T^{1-\kappa}\left|\widehat{N}_{b}-N_{b}^{0}\right|>K\left\Vert \delta^{0}\right\Vert ^{-2}\overline{\sigma}^{2}\right)<\varepsilon$;
and (iii) for $\kappa\in(0,\,1/4],$ as $T\rightarrow\infty,$ $\left(\sqrt{T/N}\left(\widehat{\beta}-\beta^{0}\right),\,\sqrt{T/N}\left(\widehat{\delta}-\delta_{h}\right)\right)'=O_{\mathbb{P}}\left(1\right)$. 
\end{prop}
The derivation of the continuous record asymptotic distribution uses
a change of time scale $s\mapsto\psi_{h}^{-1}t$ where $\psi_{h}=h^{1-\kappa}$.
Under fixed-shifts, Proposition 3.2 in CP shows that $\widehat{N}_{b}-N_{b}^{0}=O_{p}\left(T^{-1}\right)$,
i.e., $\widehat{N}_{b}$ is in a shrinking neighborhood of $N_{b}^{0}$,
which however  shrinks too fast and impedes the development of a
feasible limit theory.  Hence, the need to analyze the objective
function in a small neighborhood of the true break date under this
``fast time scale''. See Section \ref{subSection Asymptotic-Distribution-of GL LapCR}
for details. 
\begin{prop}
\label{Proposition  1 CR}Under Assumptions \ref{Assumption 1, CT}-\ref{Assumption 5 Identification}
and \ref{Assumption 6 - Small Shifts B}, and under the ``fast time
scale'', 
\begin{align}
N\left(\widehat{\lambda}_{b}-\lambda_{0}\right) & \overset{\mathcal{L}\mathrm{-}\mathrm{s}}{\Rightarrow}\underset{v\in\left[-\frac{N_{b}^{0}}{\left\Vert \delta^{0}\right\Vert ^{-2}\overline{\sigma}^{2}},\,\frac{N-N_{b}^{0}}{\left\Vert \delta^{0}\right\Vert ^{-2}\overline{\sigma}^{2}}\right]}{\mathrm{argmax}}\mathscr{V}\left(v\right),\label{CR Asymptotic Distribution}
\end{align}
where $\mathscr{V}\left(v\right)\triangleq-\left(\delta^{0}\right)'\left\langle Z_{\Delta},\,Z_{\Delta}\right\rangle \left(v\right)\delta^{0}+2\left(\delta^{0}\right)'\mathscr{W}\left(v\right),$
$\left\langle Z_{\Delta},\,Z_{\Delta}\right\rangle \left(v\right)$
is the predictable quadratic variation process of $Z_{\Delta}$ and
the process $\mathscr{W}\left(v\right)$ is, conditionally on the
$\sigma$-field $\mathscr{F}$, a two-sided centered Gaussian martingale
with independent increments.
\end{prop}
To save space, we do not report the explicit expression for the covariance
of $\mathscr{W}\left(v\right)$; see Theorem 4.1 in CP. Under stationary
regimes, Proposition \ref{Theorem 2, Asymptotic Distribution immediate Stationary Regimes}
presents the corresponding result.
\begin{assumption}
\label{Assumtpion - Regimes}The process $\Sigma^{0}$ is (possibly
time-varying) deterministic; $\left\{ z_{kh},\,e_{kh}\right\} $ is
second-order stationary within each regime. For $k=1,\ldots,\,T_{b}^{0}$,
$\mathbb{E}\left(z_{kh}z'_{kh}|\,\mathscr{F}_{\left(k-1\right)h}\right)=\Sigma_{Z,1}h$,
$\mathbb{E}\left(\widetilde{e}_{kh}^{2}|\,\mathscr{F}_{\left(k-1\right)h}\right)=\sigma_{e,1}^{2}h$
and $\mathbb{E}\left(z_{kh}z'_{kh}\widetilde{e}_{kh}^{2}|\,\mathscr{F}_{\left(k-1\right)h}\right)=\Omega_{\mathscr{W},1}h^{2}$
while for $k=T_{b}^{0}+1,\ldots,\,T$, $\mathbb{E}\left(z_{kh}z'_{kh}|\,\mathscr{F}_{\left(k-1\right)h}\right)=\Sigma_{Z,2}h$,
$\mathbb{E}\left(\widetilde{e}_{kh}^{2}|\,\mathscr{F}_{\left(k-1\right)h}\right)=\sigma_{e,2}^{2}h$
and $\mathbb{E}\left(z_{kh}z'_{kh}\widetilde{e}_{kh}^{2}|\,\mathscr{F}_{\left(k-1\right)h}\right)=\Omega_{\mathscr{W},2}h^{2}$.
\end{assumption}
Let $W_{i}^{*}\left(s\right),$ $i=1,\,2,$ be two independent standard
Wiener processes defined on $[0,\,\infty),$ starting at the origin
when $s=0.$ Let
\begin{align*}
\mathscr{V}^{*}\left(s\right) & =\begin{cases}
-\frac{\left|s\right|}{2}+W_{1}^{*}\left(s\right), & \textrm{if }s<0\\
-\frac{\left(\delta^{0}\right)'\Sigma_{Z,2}\delta^{0}}{\left(\delta^{0}\right)'\Sigma_{Z,1}\delta^{0}}\frac{\left|s\right|}{2}+\left(\frac{\left(\delta^{0}\right)'\Omega_{\mathscr{W},2}\left(\delta^{0}\right)}{\left(\delta^{0}\right)'\Omega_{\mathscr{W},1}\left(\delta^{0}\right)}\right)^{1/2}W_{2}^{*}\left(s\right), & \textrm{if }s\geq0.
\end{cases}
\end{align*}

\begin{prop}
\label{Theorem 2, Asymptotic Distribution immediate Stationary Regimes}Under
Assumptions \ref{Assumption 1, CT}-\ref{Assumption 5 Identification}
and \ref{Assumption 6 - Small Shifts B}-\ref{Assumtpion - Regimes},
and under the ``fast time scale'', 
\begin{align}
\frac{\left(\left(\delta^{0}\right)'\Sigma_{Z,1}\delta^{0}\right)^{2}}{\left(\delta^{0}\right)'\Omega_{\mathscr{W},1}\delta^{0}}N\left(\widehat{\lambda}_{b}-\lambda_{0}\right) & \Rightarrow\underset{s\in\left[-\frac{N_{b}^{0}}{\left\Vert \delta^{0}\right\Vert ^{-2}\overline{\sigma}^{2}}\frac{\left(\left(\delta^{0}\right)'\Sigma_{Z,1}\delta^{0}\right)^{2}}{\left(\delta^{0}\right)'\Omega_{\mathscr{W},1}\left(\delta^{0}\right)},\,\frac{N-N_{b}^{0}}{\left\Vert \delta^{0}\right\Vert ^{-2}\overline{\sigma}^{2}}\frac{\left(\left(\delta^{0}\right)'\Sigma_{Z,1}\delta^{0}\right)^{2}}{\left(\delta^{0}\right)'\Omega_{\mathscr{W},1}\delta^{0}}\right]}{\mathrm{argmax}}\mathscr{V}^{*}\left(s\right).\label{Equation (2) Asymptotic Distribution}
\end{align}
\end{prop}

\subsection{\label{Subsection: Comparison CR}Comparison with Other Approaches}

As shown in Figures \ref{Fig1}-\ref{Fig2}, the structural change
problem is characterized by a high degree of uncertainty when the
break magnitude is not large. The classical shrinkage asymptotics
of \citet{bai:97RES}, with $\delta_{T}$ converging to zero at a
rate slower than $O\left(T^{1/2}\right)$, underestimates the degree
of uncertainty and, as the figures show, it provides a poor approximation
to the finite-sample behavior of the LS estimator. CP argued that
this issue is also responsible for the poor coverage probabilities
of the associated confidence intervals when the break size is small.
This asymptotic distribution does not capture important features of
the finite-sample distribution such as multimodality, asymmetry and
shape changing with the magnitude of $\delta^{0}$. Underestimating
the true uncertainty leads to confidence intervals that are too short
and consequently undercover.

The goal is to find an asymptotic experiment that delivers a good
approximation and leads to inference that is reliable in practice.
Elliott and M{\"u}ller {[}cf. \citet{elliott/mueller:07} and \citet{elliott/mueller/watson:15}{]}
proposed an alternative framework where $\delta_{T}=O\left(T^{-1/2}\right)$
so that $\delta_{T}$ goes faster to zero. This can be referred to
as a weak identification. Although it increases the uncertainty in
the problem, the rate at which $\delta_{T}$ goes to zero is too fast
in the sense that statistical uncertainty is too high so that $\widehat{\lambda}_{b}=\widehat{T}_{b}/T$,
$\widehat{\delta}_{1}$ and $\widehat{\delta}_{2}$ become inconsistent
for $\lambda_{0}$, $\delta_{1}^{0}$ and $\delta_{2}^{0},$ respectively.
This can be problematic and indeed, their inference suffers from the
opposite problem in that confidence intervals for $\widehat{T}_{b}$
 can be too large and thus uninformative {[}\citet{casini/perron:OUP-Breaks},
\citet{chang/perron:18} and CP{]}. Furthermore, inconsistency for
the regression coefficients is unappealing since researchers often
are ultimately interested in making inference about the regression
coefficients and not just about $T_{b}^{0}.$

Although the asymptotic experiments considered so far all impose conditions
on the break magnitude $\delta^{0},$ CP pointed out that what matters
is not just $\delta^{0}.$ Consider a location model with a change
$\delta^{0}$ in the mean and independent errors. What describe the
uncertainty in the model is the ratio $\delta^{0}/\sigma$ where $\sigma$
is the volatility of the errors. Instead of controlling just $\delta,$
one can rather control the signal-to-noise ratio $\delta^{0}/\sigma$.
CP proposed to let $\delta^{0}$ go to zero at a not too fast rate
while letting $\sigma$ increase to infinity in a neighborhood of
$T_{b}^{0}$. That is $\left(\delta_{T}/\sigma_{t}\right)\rightarrow0$
at rate $O\left(T^{-1/2}\right)$ in a neighborhood of $T_{b}^{0}$.
This is the same rate Elliott and M{\"u}ller used for $\delta_{T}\rightarrow0.$
However, the difference here is that all the parameters in the models
remain consistent. See CP for details.

\section{Laplace-based Estimation\label{Section Asymptotic Results LapCR}}

We first formally define the Generalized Laplace (GL) estimator and
introduce assumptions needed for the derivation of its large-sample
properties. Section \ref{Subsection Asymptotic-Framework-For} describes
the asymptotic framework adopted. The main results about the limit
distribution of the estimate of $\lambda_{0}$ are presented in Section
\ref{subSection Asymptotic-Distribution-of GL LapCR}. Section \ref{subSection Feasible Method}
explains how to construct the GL estimator in practice.

\subsection{The Estimator and Additional Assumptions\label{Subsection Setting and Assumption Lap - LapCR}}

Using the properties of orthogonal LS projections, $\theta$ can be
concentrated out from the objective function $Q_{h}\left(\theta,\,N_{b}\right)$,
allowing us to dispense with the dependence of $Q_{h}\left(\theta,\,N_{b}\right)$
on $\theta$ and use $Q_{h}\left(N_{b}\right)$ hereafter. The parameter
of interest is $N_{b}$, and the Quasi-posterior density $p_{h}\left(N_{b}\right)$
is given by,
\begin{align}
p_{h}\left(N_{b}\right) & \triangleq\frac{\exp\left(Q_{h}\left(N_{b}\right)\right)\pi\left(N_{b}\right)}{\int_{\varGamma^{0}}\exp\left(Q_{h}\left(N_{b}\right)\right)\pi\left(N_{b}\right)dN_{b}},\label{Eq. Quasi-posterior}
\end{align}
which constitutes a proper distribution over the parameter space $\varGamma^{0}\triangleq\left(0,\,N\right)$.
The Quasi-prior $\pi\left(\cdot\right)$ is a weight function or simply
a prior probability density. For example, $\pi\left(N_{b}\right)=d\mathbb{Q}\left(N_{b}\right)/d\mathrm{Leb}\left(N_{b}\right)$
for some probability distribution $\mathbb{Q}$, with $\mathrm{Leb}\left(\cdot\right)$
the Lebesgue measure. The function $\pi\left(\cdot\right)$ satisfies
weak regularity conditions. Following \citet{chernozhukov/hong:03}
we restrict attention to convex loss functions $l_{h}\left(\cdot\right)$.
Common examples include, (i) $l_{h}\left(r\right)=a_{h}\left|r\right|^{m},$
the polynomial loss function with the squared loss function obtained
when $m=2$, and the absolute deviation loss with $m=1$; (ii) $l_{h}\left(r\right)=a_{h}\left(\tau-\mathbf{1}\left(r\leq0\right)\right)r,$
the check loss function where $\left\{ a_{h}\right\} $ is a positive
sequence with $a_{h}\rightarrow\infty$. Given the Quasi-posterior
density, we can define the expected risk under $p_{h}\left(\cdot\right)$,
for the loss $l_{h}\left(\cdot\right)$, as $\mathcal{R}_{l,h}\left(s\right)\triangleq\mathbb{E}_{p_{h}}\left[l_{h}\left(s-\widetilde{N}_{b}\right)\right],$
where $\widetilde{N}_{b}$ is a random variable with distribution
$p_{h}$ and $\mathbb{E}_{p_{h}}$ denotes expectation taken under
$p_{h}.$ Then, 
\begin{align}
\mathcal{R}_{l,h}\left(s\right) & =\int_{\varGamma^{0}}l_{h}\left(s-N_{b}\right)\left(\frac{\exp\left(Q_{h}\left(N_{b}\right)\right)\pi\left(N_{b}\right)}{\int_{\varGamma^{0}}\exp\left(Q_{h}\left(N_{b}\right)\right)\pi\left(N_{b}\right)dN_{b}}\right)dN_{b}.\label{Eq. Expected Risk function}
\end{align}
 The GL estimator $\widehat{N}_{b}^{\mathrm{GL}}$ is a decision rule
that is least unfavorable given the information provided by the Quasi-posterior
$p_{h}$ according to the loss function $l_{h}\left(\cdot\right)$.
That is, $\widehat{N}_{b}^{\mathrm{GL}}$ minimizes the expected risk
function in \eqref{Eq. Expected Risk function}: $\widehat{N}_{b}^{\mathrm{GL}}\triangleq\mathrm{argmin}{}_{s\in\varGamma^{0}}\mathcal{R}_{l,h}\left(s\right).$
Observe that if $p_{h}\left(N_{b}\right)$ were a true posterior,
then $\widehat{N}_{b}^{\mathrm{GL}}$ would naturally be viewed as
a Bayesian estimator for the loss function $l_{h}\left(\cdot\right)$
and prior $\pi.$ This suggests an interpretation of the Laplace-type
estimator as a Quasi-Bayesian estimator. 

In our setting, one can treat the density of the continuous record
limit distribution of $\widehat{N}_{b}^{\mathrm{LS}}$ as the ``Quasi-prior''
$\pi$. As discussed below, the resulting Quasi-posterior in \eqref{Eq. Quasi-posterior}
provides useful information for inference about the parameter $N_{b}^{0}$
beyond that already included in the  objective function $Q_{h}\left(N_{b}\right)$.
Inference procedures based on the median of the Quasi-posterior density
obtained when using the absolute loss function is a reasonable choice
as we shall show. The GL estimation provides alternative inference
methods that may be combined with the (frequentist) continuous record
asymptotic framework, through the choice of $\pi$. Note that in
order to construct such GL estimator with the continuous record Quasi-prior,
one needs an estimate of the density of the continuous record distribution,
computed as in CP; see Section \ref{subSection Feasible Method} for
details.

To understand why the GL estimation can be useful, consider Figure
\ref{Fig1}, which shows that when the magnitude of the break is small
both the finite-sample and continuous record distribution display
highly non-standard features. First, there are three modes; two are
near the start and end of sample while the mode at the origin corresponds
to the estimated break point. This multi-modality signifies that there
is a substantial tendency for the LS estimator to locate the break
date towards the tails rather than close to the true break date. Second,
the asymmetry in the density\textemdash which is always present unless
the true break date is at mid-sample\textemdash implies that the span
and actual location of the break matters for the precision of the
estimator. CP documented that such features are still present to
a lesser extent for moderate break sizes, although they disappear
when they are large (cf. Figure \ref{Fig2}). As opposed to simply
relying on $Q_{h}\left(N_{b}\right)$ which is quite flat for small
breaks, the GL estimation combines information from the continuous
record density\textemdash through $\pi\left(N_{b}\right)$\textemdash with
information from the distribution of the criterion function to yield
the Quasi-posterior. Given the highly non-standard features of the
finite-sample distribution, the Quasi-posterior contains more accurate
information about the uncertainty of the change-point. Hence, estimation
and inference based on the latter may have better properties as we
shall show.
\begin{assumption}
\label{Assumption The-loss-function LapCR}$\boldsymbol{L}$ denotes
the set of functions $l:\,\mathbb{R}\rightarrow\mathbb{R}_{+}$ that
satisfy: (i) $l\left(r\right)$ is defined on $\mathbb{R}$, with
$l\left(r\right)\geq0$ and $l\left(r\right)=0$ if and only if $r=0$;
(ii) $l\left(r\right)$ is continuous at $r=0$ but is not identically
zero; (iii) $l\left(\cdot\right)$ is convex and $l\left(r\right)\leq1+\left|r\right|^{m}$
for some $m>0$. 
\end{assumption}

\begin{assumption}
\label{Assumption Prior LapCR}The function $\pi:\,\mathbb{R}\rightarrow\mathbb{R}_{+}$
is a continuous, uniformly positive density function satisfying $\pi^{0}\triangleq\pi\left(N_{b}^{0}\right)>0,$
and $\pi^{0}<C_{\pi}$ for some finite $C_{\pi}$. Furthermore, $\pi\left(N_{b}\right)=0$
for all $N_{b}\notin\varGamma^{0}$, and $\pi$ is twice continuously
differentiable in $N_{b}$ at $N_{b}^{0}$. 
\end{assumption}
\begin{defn}
\label{Assumption Uniquness LapCR}The random variable $\xi_{l}^{0}$
is uniquely defined by
\begin{align*}
\Psi_{l}^{*}\left(\xi_{l}^{0}\right) & \triangleq\inf_{s}\Psi_{l}^{*}\left(s\right)=\inf_{s}\int_{\Gamma^{*}}l\left(s-v\right)\left(e^{\mathscr{V}\left(v\right)}\pi\left(N_{b}^{0}+v/\vartheta\right)/\int_{\Gamma^{*}}e^{\mathscr{V}\left(w\right)}\pi\left(N_{b}^{0}+w/\vartheta\right)dw\right)dv
\end{align*}
 where $\vartheta\triangleq\left\Vert \delta^{0}\right\Vert /\sigma^{2}$;
$\Gamma^{*}$ and the process $\mathscr{V}\left(\cdot\right)$ are
specified below.
\end{defn}

\begin{lem}
\label{Lemma Identification con. holds}For any $\eta>0$, there exists
an $\epsilon>0$, such that 
\begin{align}
\liminf_{h\downarrow0}\,\mathbb{P} & \left[\sup_{\left|N_{b}-N_{b}^{0}\right|\geq\eta}Q_{h}\left(N_{b}\right)-Q_{h}\left(N_{b}^{0}\right)\leq-\epsilon\right]=1.\label{Eq. Identification}
\end{align}
\end{lem}
The conditions in Assumption \ref{Assumption The-loss-function LapCR}
are similar in spirit to those in \citet{bickel/yahav:69} and \citet{chernozhukov/hong:03}.
A convex loss function is usually employed in applications. The restriction
imposed in part (iii) is not essential. What one needs is that the
growth of the function $l\left(r\right)$ as $\left|r\right|\rightarrow\infty$
is slower than that of functions of the form $\exp\left(\epsilon\left|r\right|\right),\,\epsilon>0.$
Further, note that part (iii) implicitly assumes that $l\left(\cdot\right)$
has been scaled by some constant for the inequality to hold. Assumption
\ref{Assumption Prior LapCR} on the Quasi-prior is rather mild. As
anticipated above, one may take $\pi\left(\cdot\right)$ as the density
of the limit process appearing in Proposition \ref{Proposition  1 CR}.
The uniqueness of $\xi_{l}^{0}$ follows from the fact that there
is a single break. Lemma \ref{Lemma Identification con. holds} guarantees
the identification of the parameters.

\subsection{\label{Subsection Asymptotic-Framework-For}Asymptotic Framework
for the Generalized Laplace Estimation }

Our framework builds upon an expansion of the criterion function derived
by CP under a continuous record. This is useful because it concentrates
out the regression parameters and allows us to frame our theory in
terms of a single local parameter $u=r_{h}\left(N_{b}-N_{b}^{0}\right)$,
while the regression parameters are kept fixed at their true values;
$r_{h}$ is some sample size-dependent sequence which converges to
infinity and whose properties will be specified below. This makes
$Q_{h}\left(\theta\left(N_{b}\right),\,N_{b}\right)=\delta'\left(N_{b}\right)\left(Z'_{2}M_{X}Z_{2}\right)\delta\left(N_{b}\right)$
a function only of the localized break date $u=r_{h}\left(N_{b}-N_{b}^{0}\right)$.
Because there is a single break in the model, it follows that the
unknown parameter $N_{b}^{0}\in\varGamma^{0}\subset\left(0,\,N\right)$
is the unique maximizer of $Q_{0,h}\left(\theta\left(N_{b}\right),\,N_{b}\right)=\mathbb{E}\left[Q_{h}\left(\theta\left(N_{b}\right),\,N_{b}\right)-Q_{h}\left(\theta\left(N_{b}^{0}\right),\,N_{b}^{0}\right)\right]$.
Further, let $\overline{Q}_{h}\left(\theta\left(N_{b}\right),\,N_{b}\right)\triangleq Q_{h}\left(\theta\left(N_{b}\right),\,N_{b}\right)-Q_{h}(\theta\left(N_{b}^{0}\right),$
$\,N_{b}^{0})$ and $G_{h}\left(\theta\left(N_{b}\right),\,N_{b}\right)\triangleq\overline{Q}_{h}\left(\theta\left(N_{b}\right),\,N_{b}\right)-Q_{0,h}\left(\theta\left(N_{b}\right),\,N_{b}\right).$
These expansions of the criterion function shows that $G_{h},\,\overline{Q}_{h}$
and $Q_{0,h}$ do not depend on $\theta\left(N_{b}\right)$ but only
on $\theta^{0}$. Hence, since $\theta^{0}\in\mathbf{S}\subset\mathbb{R}^{p+q}\times\mathbb{R}^{q}$
is given, we shall omit it from the arguments of $G_{h},\,\overline{Q}_{h}$
and $Q_{0,h}$ in what follows.\footnote{The exact expressions for $G_{h},\,\overline{Q}_{h}^{\mathrm{}}$
and $Q_{0,h}$ are provided at the beginning of Section A.2 of the
supplement.}\textcolor{black}{} The GL estimator $\widehat{N}_{b}^{\mathrm{GL}}$
can equivalently be defined as the minimizer of
\begin{align}
\Psi_{l,h}\left(s\right) & =\int_{\varGamma^{0}}l\left(s-N_{b}\right)\frac{\exp\left(\gamma_{h}\overline{Q}_{h}\left(N_{b}\right)\right)\pi\left(N_{b}\right)}{\int_{\varGamma^{0}}\exp\left(\gamma_{h}\overline{Q}_{h}\left(N_{b}\right)\right)\pi\left(N_{b}\right)dN_{b}}dN_{b},\label{eq. Definition GLE, Psi(s, theta) eq. 5 LapCR}
\end{align}
where $\left\{ \gamma_{h}\right\} $ is a sequence which normalizes
the sample criterion. Conditions on $\left\{ \gamma_{h}\right\} $
will be stated below. The main theoretical result of this section
concerns the large-sample properties of the GL estimator, which we
derive as follows. We first show the convergence of the marginal
distributions of the sample function $\Psi_{l,h}\left(s\right)$ to
the marginal distributions of the random function $\Psi_{l}^{*}\left(s\right)=\int_{\Gamma^{*}}l\left(s-v\right)\exp\left(\mathscr{V}\left(v\right)\right)\pi\left(N_{b}^{0}+v/\vartheta\right)/\left(\int_{\Gamma^{*}}\exp\left(\mathscr{V}\left(w\right)\right)\pi\left(N_{b}^{0}+w/\vartheta\right)dw\right)du$,
where $\pi_{h}\left(v\right)$ is defined below. Let $\mathbf{K}\triangleq\left\{ s\in\mathbb{R}:\,\left|s\right|\leq K<\infty\right\} $.
Next, we show that the family of probability measures in $\mathbb{C}_{b}\left(\mathbf{K}\right)$
(the space of bounded continuous function from $\mathbf{K}$ into
$\mathbb{R}$), generated by the contractions of $\Psi_{l,h}^{*}\left(s\right)$
on $\mathbf{K}$, is dense. As a final step, we analyze the oscillations
of the minimum points of the sample criterion $\Psi_{l,h}^{*}\left(s\right)$.
Given $u=r_{h}\left(N_{b}-N_{b}^{0}\right)$, we let $\pi_{h}\left(u\right)\triangleq\pi\left(N_{b}^{0}+u/r_{h}\right),$
$\widetilde{Q}_{h}\left(u\right)\triangleq\overline{Q}_{h}\left(N_{b}^{0}+u/r_{h}\right),$
$\widetilde{G}_{h}\left(u\right)\triangleq G_{h}\left(N_{b}^{0}+u/r_{h}\right)$
and $\widetilde{Q}_{0,h}\left(u\right)\triangleq Q_{0,h}\left(N_{b}^{0}+u/r_{h}\right)$.
Using the expression for $u,$ we can apply a simple substitution
in \eqref{eq. Definition GLE, Psi(s, theta) eq. 5 LapCR} which results
in,
\begin{align}
\Psi_{l,h}\left(s\right) & =\int_{\Gamma_{h}}l\left(s-u\right)\frac{\exp\left(\gamma_{h}\left(\widetilde{G}_{h}\left(u\right)+\widetilde{Q}_{0,h}\left(u\right)\right)\right)\pi_{h}\left(u\right)du}{\int_{\Gamma_{h}}\exp\left(\gamma_{h}\left(\widetilde{G}_{h}\left(w\right)+\widetilde{Q}_{0,h}\left(w\right)\right)\right)\pi_{h}\left(w\right)dw},\label{eq. (17)-1-1}
\end{align}
 where $\Gamma_{h}\triangleq\left\{ u\in\mathbb{R}:\,N_{b}^{0}+u/r_{h}\in\varGamma^{0}\right\} $.
 The local parameter $u$ introduced above depends on the normalizing
factor $\left\{ r_{h}\right\} $. We set $r_{h}\triangleq T^{1-\kappa}/\vartheta N$,
$\kappa\in\left(0,\,1/2\right)$ with $\vartheta\triangleq\left\Vert \delta^{0}\right\Vert ^{2}/\overline{\sigma}^{2}$
 so as to have\footnote{$N_{b}=N\lambda_{b}$ and so theoretical results about $\lambda_{b}$
translate immediately to $N_{b}$ (up to a constant $N$).} 
\begin{align}
u & \triangleq\vartheta^{-1}T^{1-\kappa}\left(\lambda_{b}-\lambda_{0}\right).\label{Definition u}
\end{align}
 The factor $T^{1-\kappa}$ is the rate at which $N\widehat{\lambda}_{b}^{\mathrm{LS}}$
convergences to $N_{b}^{0}$ under the continuous record asymptotic
setting. Note that the local parameter $u$ is allowed to vary over
the entire real line. On the range $\left\{ \left|u\right|<K\right\} $,
$N_{b}$ approaches $N_{b}^{0}$ at the rate $T^{1-\kappa}$. Thus,
we will deduce a convergence result for the normalized criterion function
$\gamma_{h}\widetilde{Q}_{h}\left(u\right)$ toward a tight Gaussian
process. For other values of $u,$ we show that the tails of the Quasi-posterior
are negligible. For the latter, one uses properties of the Gaussian
component of the limit process $\mathscr{V}\left(\cdot\right),$ and
show that it cannot diverge faster than the (negative) drift component.
Since the Quasi-posterior is an exponential transform of the centered
objective function, the tails of some expansion of the criterion function
diverge (after rescaling) to minus infinity. This follows because
the objective function can only be maximized for values of the parameter
sufficiently close to $N_{b}^{0},$ and so its re-centered version
is always negative and bounded away from zero if $N_{b}$ is far from
$N_{b}^{0}$. The normalizing sequence $\left\{ \gamma_{h}\right\} $
then makes the exponential transformation negligible for such $N_{b}$
far from $N_{b}^{0}$.

\subsection{\label{subSection Asymptotic-Distribution-of GL LapCR}Asymptotic
Distribution of the Generalized Laplace Estimator}

The main result of this section is Theorem \ref{Theorem  LapCR General},
which presents the limit distribution of the GL estimator for a general
loss function $l\left(\cdot\right)$. As part of the proof, we show
the weak convergence of $\widetilde{Q}_{h}\left(u\right)$ on the
space of bounded functions from compact sets $\mathbf{B}\subset\mathbb{R}$
into $\mathbb{R}$, denoted by $\mathbb{D}_{b}\left(\mathbf{B}\right)$.
As a matter of notation let $\widetilde{\mathscr{W}}\left(\theta,\,u\right)$
denote an arbitrary sample process with bounded \textit{}paths evaluated
at $\left(\theta,\,u\right)$ with $u\in\mathbb{R}$. For each fixed
$\theta$, we write $\widetilde{\mathscr{W}}\left(\theta,\,u\right)\overset{}{\Rightarrow}\mathscr{W}\left(\theta,\,u\right)$
on the space $\mathbb{D}_{b}\left(\mathbf{B}\right)$ if the process
$\widetilde{\mathscr{W}}\left(\theta,\,\cdot\right)$ converges 
in law under the Skorohod metric to a process $\mathscr{W}\left(\theta,\,\cdot\right)$
defined on $\mathbb{D}_{b}\left(\mathbf{B}\right)$. To simplify notation
we omit the argument $\theta$ from the limit process. For general
loss functions, the Laplace estimator is defined implicitly as the
solution of a convex optimization problem. The theorem presents
the limit distribution of the estimator under the change of time scale,
discussed above. Formally, the limiting distribution is derived under
a change of time scale $s\mapsto\psi_{h}^{-1}s$. Then, $t\triangleq\psi_{h}^{-1}s$
is the index on the new time scale. Hence, the sample criterion function
$\widetilde{Q}_{h}\left(v\right)=\widetilde{G}_{h}\left(v\right)+\widetilde{Q}_{0,h}\left(v\right)$
will be shown to vary on $\Gamma^{*}\triangleq\left\{ v\in\mathbb{R}:\,-\vartheta N_{b}^{0}\leq v\leq\vartheta\left(N-N_{b}^{0}\right)\right\} $
where $\vartheta=\left\Vert \delta^{0}\right\Vert ^{2}\overline{\sigma}^{-2}$
and the optimization problem is then also defined with respect to
$\Gamma^{*}$, namely via, 
\begin{align}
\Psi_{l,h}^{*}\left(s\right) & \triangleq\int_{\Gamma^{*}}l\left(s-u\right)\frac{\exp\left(\gamma_{h}\left(\widetilde{Q}_{h}\left(v\right)\right)\right)\pi_{h}\left(v\right)dv}{\int_{\Gamma^{*}}\exp\left(\gamma_{h}\left(\widetilde{Q}_{h}\left(w\right)\right)\right)\pi_{h}\left(w\right)dw}.\label{eq. Definition Phi*}
\end{align}
We next impose conditions on $\zeta_{k}\triangleq h^{-1}z_{kh}e_{kh}$
and $\gamma_{h}$ to derive the required limit distribution.
\begin{assumption}
\label{Assumption A.9b Bai 97, LapCR}Uniformly in $r\in\left[0,\,1\right],$
$\left(T_{b}^{0}\right)^{-1/2}\sum_{k=1}^{\left\lfloor rT_{b}^{0}\right\rfloor }\zeta_{k}\Rightarrow\mathscr{W}_{1}\left(r\right),$
$\left(T-T_{b}^{0}\right)^{-1/2}\sum_{k=T_{b}^{0}+1}^{T_{b}^{0}+\left\lfloor r\left(T-T_{b}^{0}\right)\right\rfloor }$
$\,$ $\zeta_{k}\Rightarrow\mathscr{W}_{2}\left(r\right)$  where
$\mathscr{W}_{i}\left(\cdot\right)$ is a multivariate Gaussian process
on $\left[0,\,1\right]$ with zero mean and covariance $\mathbb{E}\left[\mathscr{W}_{i}\left(u\right),\,\mathscr{W}_{i}\left(s\right)\right]=\min\left\{ u,\,s\right\} \Omega_{\mathscr{W},i}$
$\left(i=1,\,2\right)$, $\Omega_{\mathscr{W},1}\triangleq\lim_{T\rightarrow\infty}\mathbb{E}[\left(T_{b}^{0}\right)^{-1/2}\sum_{k=1}^{T_{b}^{0}}\zeta_{k}]^{2}$
and $\Omega_{\mathscr{W},2}\triangleq\lim_{T\rightarrow\infty}\mathbb{E}[\left(T-T_{b}^{0}\right)^{-1/2}\sum_{k=T_{b}^{0}+1}^{T}\zeta_{k}]^{2}$.
For any $0<r_{0}<1$ with $r_{0}\neq\lambda_{0}$, $T^{-1}\sum_{k=\left\lfloor r_{0}T\right\rfloor +1}^{\left\lfloor \lambda_{0}T\right\rfloor }h^{-1}z_{kh}z'_{kh}\overset{\mathbb{P}}{\rightarrow}\left(\lambda_{0}-r_{0}\right)\Sigma_{Z,1},$
and $T^{-1}\sum_{k=\left\lfloor \lambda_{0}T\right\rfloor +1}^{\left\lfloor r_{0}T\right\rfloor }h^{-1}z_{kh}z'_{kh}\overset{\mathbb{P}}{\rightarrow}\left(r_{0}-\lambda_{0}\right)\Sigma_{Z,2}$
with $\lambda_{-}$ and $\lambda_{+}$, the minimum and maximum eigenvalues
of the last two matrices with $0<\lambda_{-}\leq\lambda_{+}<\infty.$ 
\end{assumption}
\begin{condition}
\label{Condition 2 LapCR}As $T\rightarrow\infty$, $\gamma_{h}/T^{3/2-\kappa}\rightarrow\kappa_{\gamma}$
where $\kappa_{\gamma}>0$ is some constant.
\end{condition}
\begin{thm}
\label{Theorem  LapCR General}Let $l\in\boldsymbol{L}$ and Assumptions
\ref{Assumption Continuous Sample Path LapCR}-\ref{Assumption 5 Identification},
\ref{Assumption 6 - Small Shifts B}-\ref{Assumtpion - Regimes},
\ref{Assumption The-loss-function LapCR}-\ref{Assumption A.9b Bai 97, LapCR}
as well as Condition \ref{Condition 2 LapCR} hold. Then,  under
the ``fast time scale'' as $T\rightarrow\infty$, we have $N\left(\widehat{\lambda}_{b}^{\mathrm{GL}}-\lambda_{0}\right)\Rightarrow\xi_{l}^{0},$
where $\xi_{l}^{0}$ is defined in Definition \ref{Assumption Uniquness LapCR}
with $\Gamma^{*}=\left(-\vartheta N_{b}^{0},\,\vartheta\left(N-N_{b}^{0}\right)\right)$
and $\mathscr{V}$ from Proposition \ref{CR Asymptotic Distribution}.
\end{thm}
\begin{cor}
\label{Corollary Posterior LapCR}Under the squared loss function
$l_{h}\left(r\right)=a_{h}\left|r\right|^{2}$, 
\begin{align*}
N\left(\widehat{\lambda}_{b}^{\mathrm{GL}}-\lambda_{0}\right) & \Rightarrow\int_{\Gamma^{*}}v\frac{\exp\left(\mathscr{V}\left(v\right)\right)\pi\left(N_{b}^{0}+v/\vartheta\right)}{\int_{\Gamma^{*}}\exp\left(\mathscr{V}\left(w\right)\right)\pi\left(N_{b}^{0}+w/\vartheta\right)dw}dv,
\end{align*}
in $\mathbb{D}_{b}\left(\mathbb{R}\right)$ with $\mathscr{V}\left(v\right)$
defined in Proposition \ref{CR Asymptotic Distribution}. Under a
least-absolute loss function, $N\left(\widehat{\lambda}_{b}^{\mathrm{GL}}-\lambda_{0}\right)$
converges to the median of $\exp\left(\mathscr{V}\left(v\right)\right)\pi\left(N_{b}^{0}+v/\vartheta\right)/\int_{\Gamma^{*}}\exp\left(\mathscr{V}\left(w\right)\right)\pi\left(N_{b}^{0}+w/\vartheta\right)dw$.
\end{cor}
Let $\rho\triangleq\left(\left(\delta^{0}\right)'\left\langle Z,\,Z\right\rangle _{1}\delta^{0}\right)^{2}/\left(\left(\delta^{0}\right)'\Omega_{\mathscr{W},1}\delta^{0}\right).$
After applying the usual change in variables {[}cf. \citet{bai:97RES},
\citet{bai/perron:98} and CP{]}, we obtain a limit distribution expressed
directly in terms of quantities that can be estimated. 
\begin{cor}
\label{Corollary Limti Distrbu}Let $l\in\boldsymbol{L}$ and Assumptions
\ref{Assumption Continuous Sample Path LapCR}-\ref{Assumption 5 Identification},
\ref{Assumption 6 - Small Shifts B}-\ref{Assumtpion - Regimes},
\ref{Assumption The-loss-function LapCR}-\ref{Assumption A.9b Bai 97, LapCR}
and Condition \ref{Condition 2 LapCR} hold. Then, under the ``fast
time scale'' and as $T\rightarrow\infty$, $N(\widehat{\lambda}_{b}^{\mathrm{GL}}-\lambda_{0})\Rightarrow\xi_{l}^{0},$
with $\xi_{l}^{0}$ uniquely defined by
\begin{align*}
\Psi_{l}^{*}\left(\xi_{l}^{0}\right) & =\inf_{s}\Psi_{l}^{*}\left(s\right)=\inf_{s}\int_{\Gamma_{\rho}^{*}}l\left(s-v\right)\frac{\exp\left(\mathscr{V}^{*}\left(v\right)\right)\pi\left(N_{b}^{0}+v/\vartheta\rho\right)}{\int_{\Gamma_{\rho}^{*}}\exp\left(\mathscr{V}^{*}\left(w\right)\right)\pi\left(N_{b}^{0}+w/\vartheta\rho\right)dw}dv
\end{align*}
 where $\Gamma_{\rho}^{*}=\left(-\vartheta\rho N_{b}^{0},\,\vartheta\rho\left(N-N_{b}^{0}\right)\right)$
with $\mathscr{V}^{*}\left(\cdot\right)$ defined in Proposition \ref{Theorem 2, Asymptotic Distribution immediate Stationary Regimes}.
\end{cor}
Theorem \ref{Theorem  LapCR General} states that the asymptotic distribution
of the GL estimator\textemdash under the new ``fast time scale''
asymptotic framework\textemdash is an integral-ratio of functions
of Gaussian processes. An advantage is that the knowledge of $\kappa$,
which determines the rate of convergence on the original time scale,
is not needed to conduct inference. 

We can compare this limiting distribution with that of the Bayesian
change-point estimator of \citet{ibragimov/has:81} {[}see their equation
(2.17) on p. 338{]}. They considered maximum likelihood and Bayesian
estimators of the change-point in a simple diffusion process. We note
a few differences. First, the region of integration is $\Gamma^{*}$
instead of $\mathbb{R}$. Second, due to the change of time scale,
the Quasi-prior $\pi\left(\cdot\right)$ enters the limiting distribution,
a useful property in view of the features of the finite-sample distributions.
It reflects the fact that the uncertainty in a change-point problem
is often high enough that the information provided by the prior can
influence the limiting behavior. Finally, note that the GL estimator
conserves a classical (frequentist) interpretation. 

\subsection{\label{subSection Feasible Method}Construction of the GL Estimate}

The definition of the GL estimate involves a component, $\exp\left(Q_{h}\left(N_{b}\right)\right)/\int_{\varGamma^{0}}\exp\left(Q_{h}\left(N_{b}\right)\right)dN_{b}$,
which is immediately available while the Quasi-prior $\pi\left(N_{b}\right)$
needs to be replaced by a consistent estimate. This requires to obtain
by simulations an estimate of the density of the continuous record
limiting distribution in \eqref{Equation (2) Asymptotic Distribution}.
We follow CP. Theorem \ref{Proposition  1 CR} shows that the limiting
distribution of the LS break point estimator is related to the distribution
of the location of the maximum of the process $\mathscr{V}\left(s\right)=\mathscr{W}\left(s\right)-\varLambda\left(s\right)$,
or after a change in variable, of the process, 
\begin{align}
\mathscr{V}^{*}\left(s\right) & =\begin{cases}
-\frac{\left|s\right|}{2}+W_{1}^{*}\left(-s\right), & \textrm{if }s<0\\
-\frac{s}{2}\phi_{Z}+\phi_{e}W_{2}^{*}\left(s\right), & \textrm{if }s\geq0,
\end{cases}\label{eq. V(s) Inference}
\end{align}
where $W_{i}^{*}\,\left(i=1,\,2\right)$ are independent standard
Wiener processes, $\phi_{Z}\triangleq\left(\delta^{0}\right)'\Sigma_{Z,2}\delta^{0}/\left(\delta^{0}\right)'\Sigma_{Z,1}\delta^{0}$
and $\phi_{e}\triangleq\left(\delta^{0}\right)'\Omega_{\mathscr{W},2}\delta^{0}/\left(\delta^{0}\right)'\Omega_{\mathscr{W},1}\delta^{0}$.
Further, for every $t,\,s\in\mathbb{R}_{+}$, let $\varSigma^{0}\left(t,\,s\right)\triangleq\mathbb{E}\left(\mathscr{W}\left(t\right),\,\mathscr{W}\left(s\right)\right).$
To conduct inference, one needs estimates for $N_{b}^{0}$, $\rho,$
$\rho\vartheta N_{b}^{0}$, $\phi_{Z}$ and $\phi_{e}$. With the
normalization $N=1$, $\widehat{\lambda}_{b}^{\mathrm{LS}}=\widehat{T}_{b}^{\mathrm{LS}}/T$
is a natural estimate of $\lambda_{0}$. Consistent estimates of $\phi_{Z}$
and $\phi_{e}$ are given by 
\begin{align*}
\widehat{\phi}_{Z} & =\frac{\widehat{\delta}'\left(T-\widehat{T}_{b}^{\mathrm{LS}}\right)^{-1}\sum_{k=\widehat{T}_{b}^{\mathrm{LS}}+1}^{T}z_{kh}z'_{kh}\widehat{\delta}}{\widehat{\delta}'\left(\widehat{T}_{b}^{\mathrm{LS}}\right)^{-1}\sum_{k=1}^{\widehat{T}_{b}^{\mathrm{LS}}}z_{kh}z'_{kh}\widehat{\delta}}\\
\widehat{\phi}_{e} & =\frac{\widehat{\delta}'\left(T-\widehat{T}_{b}^{\mathrm{LS}}\right)^{-1}\sum_{k=\widehat{T}_{b}^{\mathrm{LS}}+1}^{T}\widehat{e}_{kh}^{2}z_{kh}z'_{kh}\widehat{\delta}}{\widehat{\delta}'\left(\widehat{T}_{b}^{\mathrm{LS}}\right)^{-1}\sum_{k=1}^{\widehat{T}_{b}^{\mathrm{LS}}}\widehat{e}_{kh}^{2}z_{kh}z'_{kh}\widehat{\delta}},
\end{align*}
 where $\widehat{\delta}$ and $\widehat{e}_{kh}$ are the LS estimator
of $\delta_{h}$ and the residuals. Let $\vartheta=\left\Vert \delta^{0}\right\Vert ^{2}\overline{\sigma}^{-2}\rho$
and

\begin{align*}
\widehat{\vartheta}= & \widehat{\rho}\left\Vert \widehat{\delta}\right\Vert ^{2}\left(T^{-1}\sum_{k=1}^{T}\widehat{e}_{kh}^{2}\right)^{-1}\frac{\left(\widehat{\delta}'\left(\widehat{T}_{b}^{\mathrm{LS}}\right)^{-1}\sum_{k=1}^{\widehat{T}_{b}^{\mathrm{LS}}}z_{kh}z'_{kh}\widehat{\delta}\right)^{2}}{\widehat{\delta}'\left(\widehat{T}_{b}^{\mathrm{LS}}\right)^{-1}\sum_{k=1}^{\widehat{T}_{b}^{\mathrm{LS}}}\widehat{e}_{kh}^{2}z_{kh}z'_{kh}\widehat{\delta}}\\
\widehat{\rho}= & \frac{\left(\widehat{\delta}'\left(\widehat{T}_{b}^{\mathrm{LS}}\right)^{-1}\sum_{k=1}^{\widehat{T}_{b}^{\mathrm{LS}}}z_{kh}z'_{kh}\widehat{\delta}\right)^{2}}{\widehat{\delta}'\left(\widehat{T}_{b}^{\mathrm{LS}}\right)^{-1}\sum_{k=1}^{\widehat{T}_{b}^{\mathrm{LS}}}\widehat{e}_{kh}^{2}z_{kh}z'_{kh}\widehat{\delta}}.
\end{align*}
 We have $\widehat{\vartheta}/h\overset{p}{\rightarrow}\vartheta$
and $\widehat{\rho}/h\overset{p}{\rightarrow}\rho$.  The final step
is to derive numerically the empirical counterpart of \eqref{eq. V(s) Inference},
i.e., to estimate the density of the continuous record limiting distribution
(cf. Proposition 5.1 in CP). The estimate is consistent under the
conditions of Theorem \ref{Theorem  LapCR General} and under fixed
shifts.

\section{\label{sec:Small-Sample-Properties-of}Small-Sample Properties of
the GL Estimators}

We conduct a Monte Carlo study to assess the small-sample accuracy
of the GL estimator. We consider a least-absolute loss function with
a continuous record prior (GL-CR) as well as a method based on the
following iterative procedure which exploits Theorem \ref{Theorem  LapCR General}.
This variant, labeled GL-CR-Iter, uses the median of the feasible
density of the continuous record distribution evaluated at the GL-CR
estimate instead of at $\widehat{N}_{b}^{\mathrm{LS}}$. Its justification
is as follows. First, when the break size is small, the objective
function is quite flat and highly variable. Hence, taking the median
of the Quasi-prior instead of $p_{h}\left(N_{b}\right)$ is likely
to involve less variability and more precise estimates. When the break
size is large, $p_{h}\left(N_{b}\right)$ and the continuous record
Quasi-prior have a similar shape with a peak at the same estimate.
In principle, this iterative procedure may be based on any estimate
$\widehat{N}_{b}$ such that $\widehat{N}_{b}=N_{b}^{0}+o_{p}\left(1\right).$
Furthermore, it can be viewed as an iterative version of the GL estimator
as defined by \eqref{eq. Definition GLE, Psi(s, theta) eq. 5 LapCR}
with $\gamma_{h}\rightarrow0$. We also consider the least-squares
(OLS) and the GL estimator under a least-absolute loss function with
a uniform prior (GL-Uni). The least-squares estimator relies on a
trimming parameter $\epsilon$. Common choices in applied work are
$\epsilon=0.15,\,0.10$ and $0.05$, which preclude locating the break
in the first and last $100\epsilon\%$ of the sample. Since the GL
estimator uses the least-squares criterion function, it also involves
the same trimming. Moreover, since it takes into account its whole
distribution, the trimming parameter plays a relatively more important
role. We recommend to set $\epsilon$ not too high, otherwise the
GL estimator tends to be too much concentrated toward the middle of
the sample for small breaks. Thus, we set $\epsilon=0.05.$ We compare
the mean absolute error (MAE), standard deviation (Std), root-mean-squared
error (RMSE), and the 25\% and 75\% quantiles.

We consider discrete-time DGPs which take the following form:
\begin{align}
y_{t}=D{}_{t}\varrho^{0}+Z{}_{t}\delta_{1}^{0}+Z{}_{t}\delta^{0}\boldsymbol{1}_{\left\{ t>T_{b}^{0}\right\} }+e_{t}, & \qquad\qquad t=1,\ldots,\,T,\label{Eq. DGP Simulation Study}
\end{align}
with $T=100.$ Three versions of \eqref{Eq. DGP Simulation Study}
are investigated: M1 is a partial structural change model with $\left\{ Z_{t}\right\} $
a zero-mean stationary Gaussian AR(1) with autoregressive coefficient
0.3 and unit innovation variance, $D_{t}=1$ for all $t$, $\varrho^{0}=1$
and $\left\{ e_{t}\right\} $ $i.i.d.$ $\mathscr{N}\left(0,\,1.21\right)$
disturbances independent of $\left\{ Z_{t}\right\} $; M2 involves
a break in the mean which corresponds to $Z_{t}=1$ for all $t$,
$D_{t}$ absent, and zero-mean stationary Gaussian AR(1) disturbances
$\left\{ e_{t}\right\} $ with AR coefficient 0.6 and innovation variance
0.49; M3 is a model with a lagged dependent variable with $D_{t}=y_{t-1}$,
$Z_{t}=1$, $e_{t}\sim i.i.d.\,\mathscr{N}\left(0,\,0.5\right)$,
$\varrho^{0}=0.6$ and $Z_{t}\delta^{0}\boldsymbol{1}_{\left\{ t>T_{b}^{0}\right\} }$
replaced by $Z_{t}(1.4\varrho^{0}\delta^{0}\boldsymbol{1}_{\left\{ t>T_{b}^{0}\right\} })$.
We set $\delta_{1}^{0}=1$ for all DGPs except in M3 where $\delta_{1}^{0}=0$.
We consider $\lambda_{0}=0.3$ and 0.5, and $\delta^{0}=0.3,\,0.4,\,0.6$
and $1$. By symmetry, the case $\lambda_{0}=0.7$ is omitted to avoid
repetitions.

The results are presented in Tables \ref{Table M3 Bias}-\ref{Table M5 Bias}.
 We document that the LS estimator displays a large absolute bias
(large MAE) when the size of the break is small,  which increases
as the true break point moves away from mid-sample. The GL estimators
successfully reduce the absolute bias; when the break point is about
mid-sample the reduction in MAE is roughly 50\% when the size of the
break is small or moderate. It is interesting to note that the distributions
of GL-CR and GL-CR-Iter are more concentrated around mid-sample than
the distribution of the OLS estimator which locates nontrivial mass
in the tails. This explains the large reduction in MAE when the break
date is about mid-sample. When the break date is close to the tails
the GL estimators still perform better than OLS, though the margin
is smaller. This feature arises because for small breaks the objective
function is quite flat with a small peak at the least-squares estimate
while the resulting Quasi-posterior inherits some trimodality from
the continuous record asymptotic distribution. Since higher mass is
located close to the least-squares estimate\textemdash which corresponds
to the middle mode\textemdash and less in the tails, the GL estimator
tends to concentrate toward mid-sample. Therefore, it is useful to
consider choosing a smaller trimming parameter so as not to miss the
information contained in the tails of the distribution and to avoid
the GL concentrating too much toward mid-sample. The smaller is $\epsilon$,
the greater is the information used by the GL method to locate the
break date.  When the size of the break is large (i.e., $\delta^{0}=1$,
bottom panel), although the absolute bias of the LS estimator becomes
relatively small, the GL estimators still have lower MAE. Notably,
the GL-type estimators display smaller variances than the LS estimator
which leads to smaller RMSE. Thus, the GL estimate based on the
CR prior dominates the LS estimator in both MAE and RMSE sense. Finally,
the GL estimator that does not use the continuous record prior (i.e.,
GL-Uni), has a performance similar to the LS estimator since the Quasi-posterior
contains essentially the same information about the objective function.
Hence, using the continuous record prior is indeed important. This
suggests that our asymptotic theory in Theorem \ref{Theorem  LapCR General}
provides a useful approximation since it states that the prior adds
useful information even in the limit. Comparing GL-CR and GL-CR-Iter,
we note that their performance is similar, though the former often
seems to be slightly more precise especially when the break is near
mid-sample.

We now discuss how the GL estimator achieves more precision relative
to LS. As shown in Figures \ref{Fig1}-\ref{Fig2} the finite-sample
distribution of the LS estimator displays trimodality when the magnitude
of the break is small. This occurs because when the evidence for a
break is weak, the LS estimator has a tendency to locate the break
in the tails as confirmed by the quantiles which show for example
that the 75\% quantile for the LS estimator is much larger than for
the GL estimators. This is a quite undesirable property given that
the break is assumed to be, e.g., in the middle 70\% of the sample
if $\epsilon=0.15$. The GL estimator does not share this property
because it tends to concentrate more mass in the middle 70\% of the
sample. This works better when the true break date is about mid-sample.
When it is near the tails, however, the choice $\epsilon=0.05$ avoids
the distribution being too concentrated near mid-sample and be more
evenly spread.  What is important is that as $\delta^{0}$ increases
we should expect the empirical distribution to move toward\textemdash and
eventually concentrates about\textemdash the true break date. This
is indeed achieved by the GL estimators.

\section{Inference Methods based on the GL Estimate\label{Section Inference-Methods-based On Laplace}}

In this section, we show how one can use the asymptotic properties
of the GL estimator in order to extend the inference procedures for
the break date presented in CP. We discuss several methods relying
on the concept of highest density region. We begin with a description
of this concept in Section \ref{subsec:Highest-Density-Regions}.
We present the new methods in Section \ref{Subsection Confidence-Sets-based}.
Section \ref{subSection: Theoretical-Results-on} provides the relevant
theory.

\subsection{\label{subsec:Highest-Density-Regions}Highest Density Regions}

CP proposed to construct confidence sets using the concept of highest
density region (HDR). The limit probability distribution of the break
date estimator  is given by \eqref{Equation (2) Asymptotic Distribution},
which is non-pivotal and depends on nuisance parameters. However,
it can be simulated as described in Section \ref{subSection Feasible Method}.
The next step is to derive numerically the regions of high density
which can be used to construct the confidence sets for $T_{b}^{0}$.
The method based on HDR is especially useful when the distribution
of the estimate is multi-modal and/or asymmetric.  When the density
is symmetric and uni-modal, the HDR method reduces to the conventional
one to construct confidence sets, i.e., symmetric and based on standard
errors. CP discussed an algorithm to obtain the HDR following \citet{hyndman:96}
{[}for more recent developments see \citet{samworth/wand:10} and
\citeauthor{mason/polonik:09} (2008, 2009){]}.\nocite{mason/polonik:08}
Choose some $0<\alpha<1$ and let $\widehat{\mathbb{P}}_{T_{b}}^{\mathrm{CR}}$
denote the empirical counterpart of the limit distribution in Proposition
\ref{Theorem 2, Asymptotic Distribution immediate Stationary Regimes}
and $\widehat{\mu}^{\mathrm{CR}}$ its corresponding density function.
\begin{defn}
\label{Def Highest-Density-Region}\textbf{Highest Density Region}:
Assume that the density function $f_{Y}\left(y\right)$ of a random
variable $Y$ defined on a probability space $\left(\Omega_{Y},\,\mathscr{F}_{Y},\,\mathbb{P}_{Y}\right)$
and taking values on the measurable space $\left(\mathcal{Y},\,\mathscr{Y}\right)$
is continuous and bounded. The $\left(1-\alpha\right)100\%$ HDR is
a subset $\mathbf{S}\left(\kappa_{\alpha}\right)$ of $\mathcal{Y}$
defined as $\mathbf{S}\left(\kappa_{\alpha}\right)=\{y:\,f_{Y}\left(y\right)>\kappa_{\alpha}\}$
where $\kappa_{\alpha}$ is the largest constant that satisfies $\mathbb{P}_{Y}\left(Y\in\mathbf{S}\left(\kappa_{\alpha}\right)\right)\geq1-\alpha$. 
\end{defn}

\begin{defn}
\label{Def. Confidence Sets for Break Date-1}\textbf{OLS-CR HDR-based
Confidence Sets for $T_{b}^{0}$}: A $\left(1-\alpha\right)100\%$
HDR-based confidence set for $T_{b}^{0}$ is a subset of $\left\{ 1,\ldots,\,T\right\} $
given by $C(\textrm{cv}_{\alpha})=\{T_{b}\in\{1,\ldots,\,T\}:\,T_{b}\in\mathbf{S}(\mathrm{cv}_{\alpha})\},$
where $\mathbf{S}\left(\textrm{cv}_{\alpha}\right)=\{T_{b}:\,\widehat{\mu}^{\mathrm{CR}}>\mathrm{cv}_{\alpha}\}$
with $\mathrm{cv}_{\alpha}$ satisfying $\sup_{\mathrm{cv}_{\alpha}\in\mathbb{R}_{+}}$
$\widehat{\mathbb{P}}_{T_{b}}^{\mathrm{CR}}(T_{b}\in\mathbf{S}(\mathrm{cv}_{\alpha}))\geq1-\alpha$.
\end{defn}
The confidence set $C\left(\mathrm{cv}_{\alpha}\right)$ has a frequentist
interpretation even though the concept of HDR is often encountered
in Bayesian analyses since it associates naturally to the derived
posterior distribution, especially when the latter is multi-modal.
Another feature of the confidence set $C\left(\mathrm{cv}_{\alpha}\right)$
is that it may consist of the union of several disjoint intervals
when the size of the shift is small.  In summary, one needs to carry
out the following steps to construct the HDR-based confidence sets
when using the LS estimate, labeled OLS-CR.
\begin{lyxalgorithm}
\label{Alg: HDR-based-Confidence-sets}\textbf{OLS-CR HDR-based Confidence
sets for $T_{b}^{0}$}\\
(1) Estimate by LS the date of the break and the regression coefficients
from model \eqref{Model Matrix format, CT}; (2) Replace the break
date $N_{b}^{0}$ by $\widehat{N}_{b}^{\mathrm{LS}},$ and the other
quantities appearing in \eqref{Equation (2) Asymptotic Distribution}
by corresponding estimates as explained in Section \ref{subSection Feasible Method};
(3) Simulate the limit distribution $\widehat{\mathbb{P}}_{T_{b}}^{\mathrm{CR}}$
from \eqref{Equation (2) Asymptotic Distribution}; (4) Compute the
HDR of the empirical distribution $\widehat{\mathbb{P}}_{T_{b}}^{\mathrm{CR}}$
and include the point $T_{b}$ in the level $\left(1-\alpha\right)$
confidence set $C\left(\mathrm{cv}_{\alpha}\right)$ if $T_{b}$ satisfies
the conditions in Definition \ref{Def. Confidence Sets for Break Date-1}.

\end{lyxalgorithm}

\subsection{\label{Subsection Confidence-Sets-based}Confidence Sets based on
$\widehat{N}_{b}^{\mathrm{GL}}$ with $\pi\left(N_{b}\right)=\widehat{\mu}^{\mathrm{CR}}\left(N_{b}\right)$}

We use the definition in \eqref{Eq. Quasi-posterior} with the Quasi-prior
$\pi\left(N_{b}\right)$ given by the continuous record asymptotic
density $\widehat{\mu}^{\mathrm{CR}}$.  We introduce the concept
of Highest Quasi-posterior Density (HQPD) regions, defined analogously
to HDR in Definition \ref{Def Highest-Density-Region} with $p_{h}\left(N_{b}\right)$
being the object of interest. We then construct the confidence sets
for $T_{b}^{0}$ as follows. We label this method by GL-CR.
\begin{lyxalgorithm}
\label{Alg: HQPD}\textbf{GL-CR HQPD-based Confidence sets for} $T_{b}^{0}$
\\
(1) Estimate by least-squares $N_{b}^{0}$ and the regression coefficients
from model \eqref{Model (4), scalar format, CT}; (2) Replace the
break date $N_{b}^{0}$ by $\widehat{N}_{b}^{\mathrm{LS}}$, and the
other quantities appearing in \eqref{Equation (2) Asymptotic Distribution}
by corresponding estimates as explained in Section \ref{subSection Feasible Method};
(3) Simulate the limit distribution $\widehat{\mathbb{P}}_{T_{b}}^{\mathrm{CR}}$
from \eqref{Theorem 2, Asymptotic Distribution immediate Stationary Regimes}
and set the Quasi-prior $\pi\left(N_{b}\right)$ equal to the probability
density of $\widehat{\mathbb{P}}_{T_{b}}^{\mathrm{CR}}$; (4) Construct
the Quasi-posterior \eqref{Eq. Quasi-posterior}; (5) Obtain numerically
the limit probability distribution of $\widehat{T}_{b}^{\mathrm{GL}}$
given in Corollary \ref{Corollary Limti Distrbu} and label it $\mathbb{\widehat{P}}_{T_{b}}^{\mathrm{GL}}$;
(6) Compute the HQPD region of the probability distribution $\mathbb{\widehat{P}}_{T_{b}}^{\mathrm{GL}}$
and include the point $T_{b}$ in the level $\left(1-\alpha\right)\%$
confidence set $C_{\mathrm{GL-CR}}\left(\mathrm{cv}_{\alpha}\right)$
if $T_{b}$ satisfies the conditions in Definition \ref{Def. Confidence Sets for Break Date-1}.
\end{lyxalgorithm}
The confidence set $C_{\mathrm{GL-CR}}\left(\mathrm{cv}_{\alpha}\right)$
may also be interpreted as a Quasi-Bayes confidence set. Step 5 involves
the derivation of the empirical counterpart of the limiting distribution
of the GL estimator. In step 3, we use the continuous record Quasi-prior.
Any prior $\pi\left(N_{b}\right)$ satisfying Assumption \ref{Assumption Prior LapCR}
could in principle be used, in which case the first three steps would
not be needed. As discussed in Section \ref{sec:Small-Sample-Properties-of},
Theorem \ref{Theorem  LapCR General} also offers the possibility
to construct alternative methods based on an iterative procedure as
follows. 
\begin{lyxalgorithm}
\label{Alg: Iterative HDR}\textbf{GL-CR-Iter HDR Confidence sets
for $T_{b}^{0}$}\\
(1) Carry out steps 1-3 in Algorithm \ref{Alg: HQPD}; (2) Construct
the Quasi-posterior in \eqref{Eq. Quasi-posterior} and compute the
estimate $\widehat{N}_{b}^{\mathrm{GL}}$; (3) Replace the break date
$N_{b}^{0}$ by $\widehat{N}_{b}^{\mathrm{GL}},$ and the other quantities
appearing in \eqref{Equation (2) Asymptotic Distribution} by estimates
as explained in Section \ref{subSection Feasible Method}; (4) Simulate
the empirical counterpart of the probability distribution from Proposition
\ref{Theorem 2, Asymptotic Distribution immediate Stationary Regimes}
using step 3 labeled $\widehat{\mathbb{P}}_{T_{b}}^{\mathrm{CR-GL}}$;
(5) Compute the HDR of the empirical distribution $\widehat{\mathbb{P}}_{T_{b}}^{\mathrm{CR-GL}}$
and include $T_{b}$ in the level $\left(1-\alpha\right)$ confidence
set $C_{\mathrm{GL-CR-Iter}}\left(\textrm{cv}_{\alpha}\right)$ if
it satisfies the conditions in Definition \ref{Def. Confidence Sets for Break Date-1}.
\end{lyxalgorithm}

\subsection{\label{subSection: Theoretical-Results-on}Theoretical Results on
Inference Methods }

As explained above one generally needs estimates of some population
quantities appearing in the continuous record asymptotic distribution
in \eqref{CR Asymptotic Distribution} and in the asymptotic distribution
of the GL estimator. They can be constructed as explained in Section
\ref{subSection Feasible Method}.
\begin{assumption}
\label{Assumption Laplace Inference LapCR} Let $\widehat{\varSigma}_{i,h}\left(\cdot\right)$
be based on the estimates $\widehat{N}_{b,h},\,\widehat{\delta}_{h},\,\widehat{\phi}_{Z,h},$
and $\widehat{\phi}_{e,h}$ discussed in Section \ref{subSection Feasible Method}.
For all $t,\,s\in\mathbb{R}$ and any $c>0$, $\widehat{\varSigma}_{i,h}\left(\cdot\right)$
satisfies, for $i=1,\,2$: (i) $\widehat{\varSigma}_{i,h}\left(t,\,s\right)=\varSigma_{i}^{0}\left(t,\,s\right)+o_{\mathbb{P}}\left(1\right)$;
(ii) $\widehat{\varSigma}_{i,h}\left(t,\,s\right)=\widehat{\varSigma}_{i,h}\left(t,\,t\right)$
if $t<s$ and $\widehat{\varSigma}_{i,h}\left(t,\,s\right)=\widehat{\varSigma}_{i,h}\left(s,\,s\right)$
if $t>s$; (iii) $\widehat{\varSigma}_{i,h}\left(ct,\,ct\right)=c\widehat{\varSigma}_{i,h}\left(t,\,t\right)$;
(iv) $\mathbb{E}\{\sup_{t=1}\widehat{\varSigma}_{i,h}^{2}\left(t,\,t\right)\}=O\left(1\right)$.
\end{assumption}
Assumption \ref{Assumption Laplace Inference LapCR}-(i) requires
the availability of certain consistent estimators. Let $\left\{ \widehat{\mathscr{W}}_{h}\right\} $
be a (sample-size dependent) sequence of two-sided Gaussian processes
with covariance $\widehat{\varSigma}_{h}.$ By Assumption \ref{Assumption Laplace Inference LapCR}-(i),
the limit law of $\left\{ \widehat{\mathscr{W}}_{h}\right\} $ is
the same as the law of $\mathscr{W}$. Construct the process $\widehat{\mathscr{V}}_{h}$
by replacing the population quantities in $\mathscr{V}^{*}$ and replacing
$\mathscr{W}$ by $\widehat{\mathscr{W}}_{h}$. Parts (ii)-(iv) are
technical conditions needed for the integrability of $\exp(\widehat{\mathscr{V}}_{h}(\cdot))$
and to prove  the tightness of $\left\{ \widehat{\mathscr{W}}_{h}\right\} $.
Introduce the following random sample quantity:
\begin{align*}
\widehat{\Psi}_{l,h}\left(s\right) & \triangleq\int_{\widehat{\Gamma}^{*}}l\left(s-v\right)\frac{\exp\left(\widehat{\mathscr{V}}_{h}\left(v\right)\right)\pi\left(\widehat{N}_{b}^{\mathrm{GL}}+v\right)}{\int_{\widehat{\Gamma}^{*}}\exp\left(\widehat{\mathscr{V}}_{h}\left(w\right)\right)\pi\left(\widehat{N}_{b}^{\mathrm{GL}}+w\right)dw}dv
\end{align*}
where $\widehat{\Gamma}^{*}$ uses the estimates in Assumption \ref{Assumption Laplace Inference LapCR}
instead of the true values. Further, let $\widehat{\xi}_{l,h}$ be
the absolute minimum point of $\widehat{\Psi}_{l,h}\left(s\right)$.
Most inference methods introduced in the previous sub-section involve
a numerical simulation of $\widehat{\Psi}_{l,h}$ or the moments or
functions of the Quasi-posterior. The following theorem shows that
$\widehat{\xi}_{l,h}$ converges in distribution to $\xi_{l}^{0}$.
\begin{thm}
\label{Theorem Inference Limi Distr LapCR}Let $l\in\boldsymbol{L}$.
Under Assumptions \ref{Assumption Continuous Sample Path LapCR}-\ref{Assumption 5 Identification},
\ref{Assumption The-loss-function LapCR}-\ref{Assumption A.9b Bai 97, LapCR}
and \ref{Assumption Laplace Inference LapCR}, the distribution of
$\widehat{\xi}_{l,h}$ is first-order equivalent to the distribution
of $\xi_{l}^{0}$ in Theorem \ref{Theorem  LapCR General}.
\end{thm}
By Theorem \ref{Theorem Inference Limi Distr LapCR}, $\widehat{\xi}_{l,h}$
can be used to conduct statistical inference. For example, in order
to use the GL HQDR-based method, step 5 of Algorithm \ref{Alg: HQPD}
requires the derivation of the empirical counterpart of $\xi_{l}^{0}$.
Theorem \ref{Theorem Inference Limi Distr LapCR} shows the validity
of methods based on numerically evaluating $\widehat{\xi}_{l,h}$.
This is achieved by simulating $\widehat{\Psi}_{l,h}$ using of the
estimators in Assumption \ref{Assumption Laplace Inference LapCR}.
Since $\widehat{\xi}_{l,h}\Rightarrow\xi_{l}^{0},$ inference based
on the probability distribution of $\widehat{\xi}_{l,h}$ is asymptotically
valid.

\section{Small-Sample Evaluation of the GL Confidence Sets\label{Section Monte Carlo Study}}

In this section, we compare the performance of the proposed confidence
sets about $T_{b}^{0}$ with existing methods. We consider Bai's
(1997) approach based on long-span shrinkage asymptotic arguments,
\citeauthor{elliott/mueller:07}\textquoteright s (2007) approach
based on\textcolor{red}{{} }inverting \citeauthor{nyblom:89}'s (1989)
statistic, the ILR method proposed by \citet{eo/morley:15} based
on the results in \citet{qu/perron:07} and the recent HDR method
proposed in CP based on the continuous record asymptotics with the
least-squares estimate (OLS-CR).\footnote{Recently, \citet{elliott/mueller/watson:15} proposed a new test for
structural breaks aimed at improving upon EM's approach. However,
they did not propose methods for the inversion of such test to construct
confidence intervals for the break date. Hence, we cannot evaluate
their method.} For brevity, we refer the readers to CP and \citet{chang/perron:18}
for a review and comprehensive evaluation of the first three methods.
A brief summary is as follows. The empirical coverage probabilities
of the confidence intervals obtained from Bai's (1997) method are
often below the nominal level when the size of the break is small.
This feature is not present for the other methods. \citeauthor{elliott/mueller:07}\textquoteright s
(2007) approach is by far the best one among existing methods in achieving
an exact coverage rate that is the closest to the nominal level. However,
this comes at the cost of lengths of the confidence sets which are
always substantially larger relative to the other methods. This holds
true across all break magnitudes. In addition, in models with serially
correlated errors or lagged dependent variables, the length of the
confidence set approaches the whole sample as the magnitude of the
break increases. The ILR has coverage rates often above the nominal
level and an average length significantly longer than the OLS-CR method
at least when the magnitude of the shift is small or moderate. Further,
the ILR does not work well when the errors are heteroskedastic and
the regressor display serial correlation.

The recent OLS-CR method proposed in CP was shown to provide adequate
empirical coverage probabilities over a wide range of data-generating
mechanisms and across all break sizes and/or location of the break.
The average length of the confidence sets is always shorter than that
obtained with \citeauthor{elliott/mueller:07}'s (2007) method. The
OLS-CR method delivers confidence sets with lengths only slightly
larger than Bai's (1997) and the differences get smaller as the size
of the break increases. Overall, the OLS-CR method strikes a better
balance between accurate coverage probabilities and length of the
confidence sets. The numerical analysis in this section documents
that the confidence sets derived from the GL inference are even more
precise in terms of coverage probability than those from the OLS-CR
method and the average length is always substantially shorter than
that from \citeauthor{elliott/mueller:07}'s (2007).

We consider the same DGPs as in Section \ref{sec:Small-Sample-Properties-of}.
To construct the OLS-CR method we follow the steps outlined in the
previous section (see also CP for more details on the procedure for
models with predictable processes).  For model M2, to estimate
the long-run variance we use \citeauthor{andrews:91}' (1991) method
along with \citeauthor{andrews/monahan:92}'s (1992) AR(1) pre-whitened
two-stage procedure to select the bandwidth. We consider the version
$\widehat{\textrm{U}}_{T}\left(T_{\mathrm{m}}\right).\mathrm{neq}$
of \citet{elliott/mueller:07} that allows for heteroskedastic regimes;
using the restricted version when applicable leads to similar results. 

The least-squares estimation method is employed with a trimming parameter
$\epsilon=0.15$ and we use the required degrees of freedom adjustment
for the statistic $\widehat{\textrm{U}}_{T}$ of \citet{elliott/mueller:07}.
The OLS-CR method and the methods proposed here do not involve any
trimming  and the confidence sets can potentially include any observation.
The significance level is set at $\alpha=0.05$, and the break date
occurs at $\left\lfloor T\lambda_{0}\right\rfloor $, where $\lambda_{0}=0.2,\,0.35,\,0.5.$
The results for the 95\% nominal coverage rates are presented in Tables
\ref{Table M10}-\ref{Table M6}. Each column reports the exact coverage
rate and average length for a given break size $\delta^{0}$. The
last row of each panel includes the rejection probability of a 5\%-level
sup-Wald test using the asymptotic critical value of \citet{andrews:93};
it serves as a statistical measure about the magnitude of the break.

Overall, the simulation results confirm previous findings about the
performance of existing methods. Bai's (1997) method has a coverage
rate below the nominal coverage level when the size of the break is
small. For example, in model M2 for which there is high serial correlation
in the disturbances, it fails to display a coverage rate above 90\%
even for moderate break sizes. In contrast, the method of \citet{elliott/mueller:07}
overall yields very accurate empirical coverage rates. However, the
average length of the confidence intervals is systematically much
larger than those from all other methods across all DGPs, break sizes
and break locations.\footnote{This problem is more severe when the errors are serially correlated
or the model includes lagged dependent variables.\nocite{casini_hac}
Regarding the former, this in part may be due to issues with \citeauthor{newey/west:87}
HAC-type estimators when there are structural breaks {[}see \citeauthor{casini_CR_Test_Inst_Forecast}
(2018, 2019a, 2019b), \nocite{casini_hac}\nocite{casini_diss} \citet{casini/perron:OUP-Breaks},
\citet{chang/perron:18}, \citet{crainiceanu/vogelsang:07}, \citet{deng/perron:06},
\citet{fossati:17}, \citet{juhl/xiao:09}, \citet{kim/perron:09},
\citet{martins/perron:16}, \citet{perron/yamamoto:18} and \citet{vogeslang:99}{]}.}

The OLS-CR method provides good coverage rates for all break magnitudes
and break locations. This holds even when there is high serial correlation
in the errors (cf. M2). Furthermore, the confidence sets have average
lengths significantly shorter than those from \citet{elliott/mueller:07}.
Thus, the OLS-CR method strikes overall a good balance between accurate
coverage rates and length of the confidence sets. As observed in CP,
this method tends to display an empirical coverage rate slightly below
95\% for particular DGPs when the break size is small.

For the GL-based methods, we consider GL-CR and GL-CR-Iter implemented
with the least-absolute deviation loss function. We do not report
results with a uniform prior because as documented Section \ref{sec:Small-Sample-Properties-of},
it has poor finite-sample properties with respect to MAE and RMSE
for small breaks. In general, the family of methods based on the GL
estimator display better features compared to the OLS-CR method.
GL-CR provides shorter length than GL-CR-Iter and OLS-CR, and it displays
decent coverage rates. GL-CR-Iter provides coverage rates and lengths
similar to OLS-CR, yet the coverage rates are more accurate than GL-CR.
  Thus, we find that the GL-CR-Iter is better than GL-CR if the
primary goal is accurate coverage of the confidence sets.

In summary, the simulation results suggest that the GL-CR-Iter method
using the GL estimates is reliable as it provides accurate coverage
rates close to the nominal level and average lengths of the confidence
sets shorter relative to existing methods developed under large-$N$
asymptotics.

\section{Concluding Remarks\label{Section Conclusive-Remarks}}

Building upon the continuous record asymptotic framework of \citet{casini/perron_CR_Single_Break},
we propose a Generalized Laplace (GL) procedure for the construction
of the estimates and confidence sets for the date of a structural
change. It is defined as the minimizer of the expected risk with the
expectation taken under the Quasi-posterior, where the latter is constructed
by applying a simple transformation to the least-squares criterion
function. Our motivation stems from the non-standard properties of
the finite-sample distribution of the least-squares break point estimator.
The advantage of the GL procedure is that it combines information
from the LS estimate of the break point, the objective function and
the continuous record distribution theory. The GL estimate is more
precise than the usual LS estimate, lower MAE and RMSE, especially
when the true break date is about middle sample. In order to achieve
this result, it is advisable to choose a small trimming parameter
(e.g., $\epsilon=0.05$). We also present inference methods that use
the concept of HDR; the resulting confidence sets for the break date
strike a better balance between empirical coverage rates and average
lengths of the confidence sets relative to traditional long-span methods.
Overall, among the GL procedures, we find that GL-CR provides more
precise estimates for the break date while GL-CR-Iter results in more
accurate coverage rates.

{\small{}\newpage}{\small\par}

\begin{singlespace}

\bibliographystyle{elsarticle-harv}
\bibliography{References_JoE}
\addcontentsline{toc}{section}{References}

\end{singlespace}

\clearpage{}

\newpage{}

\appendix

\section{Appendix}

\setcounter{page}{1} \renewcommand{\thepage}{F-\arabic{page}}

\setlength{\belowcaptionskip}{-60pt}

\setlength{\belowcaptionskip}{-0pt}

\medskip{}

\newpage{}

\setcounter{page}{1} \renewcommand{\thepage}{T-\arabic{page}}

\begin{table}[H]
\caption{\label{Table M3 Bias}Small-sample accuracy of the estimates of the
break point $T_{b}^{0}$ for model M1}

\begin{centering}
{\footnotesize{}}%
\begin{tabular}{ccccccc|ccccc}
\hline 
 &  & {\footnotesize{}MAE} & {\footnotesize{}Std} & {\footnotesize{}$\textrm{RMSE}$} & {\footnotesize{}$Q_{0.25}$} & \multicolumn{1}{c}{{\footnotesize{}$Q_{0.75}$}} & {\footnotesize{}MAE} & {\footnotesize{}Std} & {\footnotesize{}$\textrm{RMSE}$} & {\footnotesize{}$Q_{0.25}$} & {\footnotesize{}$Q_{0.75}$}\tabularnewline
\cline{3-12} \cline{4-12} \cline{5-12} \cline{6-12} \cline{7-12} \cline{8-12} \cline{9-12} \cline{10-12} \cline{11-12} \cline{12-12} 
 &  & \multicolumn{5}{c|}{{\footnotesize{}$\lambda_{0}=0.3$}} & \multicolumn{5}{c}{{\footnotesize{}$\lambda_{0}=0.5$}}\tabularnewline
{\footnotesize{}$\delta^{0}=0.3$} & {\footnotesize{}OLS} & {\footnotesize{}25.04} & {\footnotesize{}30.23} & {\footnotesize{}32.50} & {\footnotesize{}18} & {\footnotesize{}72} & {\footnotesize{}25.58} & {\footnotesize{}29.15} & {\footnotesize{}29.38} & {\footnotesize{}29} & {\footnotesize{}82}\tabularnewline
 & {\footnotesize{}GL-CR} & {\footnotesize{}14.20} & {\footnotesize{}10.60} & {\footnotesize{}19.06} & {\footnotesize{}29} & {\footnotesize{}57} & {\footnotesize{}4.94} & {\footnotesize{}9.82} & {\footnotesize{}9.97} & {\footnotesize{}46} & {\footnotesize{}54}\tabularnewline
 & {\footnotesize{}GL-CR-Iter} & {\footnotesize{}17.32} & {\footnotesize{}22.64} & {\footnotesize{}22.64} & {\footnotesize{}29} & {\footnotesize{}59} & {\footnotesize{}8.16} & {\footnotesize{}9.66} & {\footnotesize{}9.69} & {\footnotesize{}41} & {\footnotesize{}59}\tabularnewline
 & {\footnotesize{}GL-Uni} & {\footnotesize{}19.17} & {\footnotesize{}20.68} & {\footnotesize{}25.32} & {\footnotesize{}28} & {\footnotesize{}62} & {\footnotesize{}16.30} & {\footnotesize{}19.67} & {\footnotesize{}19.66} & {\footnotesize{}35} & {\footnotesize{}65}\tabularnewline
\cline{3-12} \cline{4-12} \cline{5-12} \cline{6-12} \cline{7-12} \cline{8-12} \cline{9-12} \cline{10-12} \cline{11-12} \cline{12-12} 
{\footnotesize{}$\delta^{0}=0.4$} & {\footnotesize{}OLS} & {\footnotesize{}21.65} & {\footnotesize{}25.82} & {\footnotesize{}28.48} & {\footnotesize{}25} & {\footnotesize{}62} & {\footnotesize{}19.85} & {\footnotesize{}24.84} & {\footnotesize{}24.85} & {\footnotesize{}35} & {\footnotesize{}70}\tabularnewline
 & {\footnotesize{}GL-CR} & {\footnotesize{}13.31} & {\footnotesize{}11.63} & {\footnotesize{}17.83} & {\footnotesize{}28} & {\footnotesize{}50} & {\footnotesize{}4.77} & {\footnotesize{}9.75} & {\footnotesize{}9.86} & {\footnotesize{}46} & {\footnotesize{}52}\tabularnewline
 & {\footnotesize{}GL-CR-Iter} & {\footnotesize{}14.93} & {\footnotesize{}10.02} & {\footnotesize{}17.97} & {\footnotesize{}30} & {\footnotesize{}56} & {\footnotesize{}7.41} & {\footnotesize{}9.13} & {\footnotesize{}9.14} & {\footnotesize{}43} & {\footnotesize{}57}\tabularnewline
 & {\footnotesize{}GL-Uni} & {\footnotesize{}16.22} & {\footnotesize{}19.94} & {\footnotesize{}22.89} & {\footnotesize{}27} & {\footnotesize{}55} & {\footnotesize{}14.46} & {\footnotesize{}18.18} & {\footnotesize{}18.17} & {\footnotesize{}37} & {\footnotesize{}63}\tabularnewline
\cline{3-12} \cline{4-12} \cline{5-12} \cline{6-12} \cline{7-12} \cline{8-12} \cline{9-12} \cline{10-12} \cline{11-12} \cline{12-12} 
{\footnotesize{}$\delta^{0}=0.6$} & {\footnotesize{}OLS} & {\footnotesize{}12.95} & {\footnotesize{}19.92} & {\footnotesize{}20.68} & {\footnotesize{}26} & {\footnotesize{}38} & {\footnotesize{}11.69} & {\footnotesize{}17.10} & {\footnotesize{}17.10} & {\footnotesize{}44} & {\footnotesize{}56}\tabularnewline
 & {\footnotesize{}GL-CR} & {\footnotesize{}12.25} & {\footnotesize{}12.51} & {\footnotesize{}15.56} & {\footnotesize{}29} & {\footnotesize{}49} & {\footnotesize{}4.46} & {\footnotesize{}8.82} & {\footnotesize{}8.64} & {\footnotesize{}49} & {\footnotesize{}51}\tabularnewline
 & {\footnotesize{}GL-CR-Iter} & {\footnotesize{}11.99} & {\footnotesize{}8.56} & {\footnotesize{}14.72} & {\footnotesize{}30} & {\footnotesize{}44} & {\footnotesize{}5.23} & {\footnotesize{}7.13} & {\footnotesize{}7.14} & {\footnotesize{}46} & {\footnotesize{}54}\tabularnewline
 & {\footnotesize{}GL-Uni} & {\footnotesize{}11.24} & {\footnotesize{}16.19} & {\footnotesize{}17.29} & {\footnotesize{}26} & {\footnotesize{}40} & {\footnotesize{}9.60} & {\footnotesize{}13.75} & {\footnotesize{}13.36} & {\footnotesize{}44} & {\footnotesize{}56}\tabularnewline
\cline{3-12} \cline{4-12} \cline{5-12} \cline{6-12} \cline{7-12} \cline{8-12} \cline{9-12} \cline{10-12} \cline{11-12} \cline{12-12} 
{\footnotesize{}$\delta^{0}=1$} & {\footnotesize{}OLS} & {\footnotesize{}9.72} & {\footnotesize{}14.91} & {\footnotesize{}14.90} & {\footnotesize{}27} & {\footnotesize{}40} & {\footnotesize{}4.89} & {\footnotesize{}8.25} & {\footnotesize{}8.24} & {\footnotesize{}48} & {\footnotesize{}52}\tabularnewline
 & {\footnotesize{}GL-CR} & {\footnotesize{}5.06} & {\footnotesize{}8.17} & {\footnotesize{}8.21} & {\footnotesize{}27} & {\footnotesize{}32} & {\footnotesize{}2.81} & {\footnotesize{}7.17} & {\footnotesize{}7.21} & {\footnotesize{}49} & {\footnotesize{}51}\tabularnewline
 & {\footnotesize{}GL-CR-Iter} & {\footnotesize{}8.41} & {\footnotesize{}4.80} & {\footnotesize{}9.67} & {\footnotesize{}29} & {\footnotesize{}39} & {\footnotesize{}3.81} & {\footnotesize{}6.31} & {\footnotesize{}6.34} & {\footnotesize{}48} & {\footnotesize{}52}\tabularnewline
 & {\footnotesize{}GL-Uni} & {\footnotesize{}4.69} & {\footnotesize{}8.35} & {\footnotesize{}8.36} & {\footnotesize{}27} & {\footnotesize{}32} & {\footnotesize{}4.51} & {\footnotesize{}7.47} & {\footnotesize{}7.47} & {\footnotesize{}48} & {\footnotesize{}52}\tabularnewline
\hline 
\end{tabular}{\footnotesize\par}
\par\end{centering}
\noindent\begin{minipage}[t]{1\columnwidth}%
{\tiny{}The model is $y_{t}=\varrho^{0}+Z_{t}\delta_{1}^{0}+Z_{t}\delta^{0}\mathbf{1}_{\left\{ t>\left\lfloor T\lambda_{0}\right\rfloor \right\} }+e_{t},\,Z_{t}=0.3Z_{t-1}+u_{t},\,u_{t}\sim i.i.d.\,\mathscr{N}\left(0,\,1\right)\,e_{t}\sim i.i.d.\,\mathscr{N}\left(0,\,1.21\right),\,T=100$.
The columns refer to Mean Absolute Error (MAE), standard deviation
(Std), Root Mean Squared Error (RMSE) and the 25\% and 75\% empirical
quantiles. OLS is the least-squares estimator; GL-CR is the GL estimator
under a least-absolute loss function with the continuous record prior;
GL-CR-Iter is the median of the density of the continuous record distribution
which uses the GL estimator in place of $\widehat{N}_{b}^{\mathrm{LS}}$;
GL-Uni is the GL estimator under a least-absolute loss function with
a uniform prior.}%
\end{minipage}
\end{table}

\begin{table}[H]
\caption{\label{Table M2 Bias}Small-sample accuracy of the estimates of the
break point $T_{b}^{0}$ for model M2}

\begin{singlespace}
\begin{centering}
{\footnotesize{}}%
\begin{tabular}{ccccccc|ccccc}
\hline 
 &  & {\footnotesize{}MAE} & {\footnotesize{}Std} & {\footnotesize{}$\textrm{RMSE}$} & {\footnotesize{}$Q_{0.25}$} & \multicolumn{1}{c}{{\footnotesize{}$Q_{0.75}$}} & {\footnotesize{}MAE} & {\footnotesize{}Std} & {\footnotesize{}$\textrm{RMSE}$} & {\footnotesize{}$Q_{0.25}$} & {\footnotesize{}$Q_{0.75}$}\tabularnewline
\cline{3-12} \cline{4-12} \cline{5-12} \cline{6-12} \cline{7-12} \cline{8-12} \cline{9-12} \cline{10-12} \cline{11-12} \cline{12-12} 
 &  & \multicolumn{5}{c|}{{\footnotesize{}$\lambda_{0}=0.3$}} & \multicolumn{5}{c}{{\footnotesize{}$\lambda_{0}=0.5$}}\tabularnewline
{\footnotesize{}$\delta^{0}=0.3$} & {\footnotesize{}OLS} & {\footnotesize{}26.84} & {\footnotesize{}28.12} & {\footnotesize{}33.00} & {\footnotesize{}21} & {\footnotesize{}76} & {\footnotesize{}23.02} & {\footnotesize{}26.86} & {\footnotesize{}26.76} & {\footnotesize{}25} & {\footnotesize{}75}\tabularnewline
 & {\footnotesize{}GL-CR} & {\footnotesize{}12.79} & {\footnotesize{}13.13} & {\footnotesize{}18.46} & {\footnotesize{}29} & {\footnotesize{}57} & {\footnotesize{}11.84} & {\footnotesize{}13.17} & {\footnotesize{}13.12} & {\footnotesize{}35} & {\footnotesize{}65}\tabularnewline
 & {\footnotesize{}GL-CR-Iter} & {\footnotesize{}14.47} & {\footnotesize{}10.29} & {\footnotesize{}20.21} & {\footnotesize{}28} & {\footnotesize{}58} & {\footnotesize{}8.76} & {\footnotesize{}10.01} & {\footnotesize{}10.24} & {\footnotesize{}41} & {\footnotesize{}59}\tabularnewline
 & {\footnotesize{}GL-Uni} & {\footnotesize{}21.78} & {\footnotesize{}21.73} & {\footnotesize{}27.71} & {\footnotesize{}28} & {\footnotesize{}66} & {\footnotesize{}17.84} & {\footnotesize{}20.90} & {\footnotesize{}20.98} & {\footnotesize{}32} & {\footnotesize{}68}\tabularnewline
{\footnotesize{}$\delta^{0}=0.4$} & {\footnotesize{}OLS} & {\footnotesize{}23.62} & {\footnotesize{}26.99} & {\footnotesize{}30.23} & {\footnotesize{}21} & {\footnotesize{}70} & {\footnotesize{}21.23} & {\footnotesize{}25.43} & {\footnotesize{}25.44} & {\footnotesize{}25} & {\footnotesize{}75}\tabularnewline
 & {\footnotesize{}GL-CR} & {\footnotesize{}16.36} & {\footnotesize{}13.86} & {\footnotesize{}21.49} & {\footnotesize{}29} & {\footnotesize{}61} & {\footnotesize{}11.56} & {\footnotesize{}11.97} & {\footnotesize{}12.25} & {\footnotesize{}36} & {\footnotesize{}64}\tabularnewline
 & {\footnotesize{}GL-CR-Iter} & {\footnotesize{}17.19} & {\footnotesize{}10.81} & {\footnotesize{}20.35} & {\footnotesize{}28} & {\footnotesize{}57} & {\footnotesize{}8.30} & {\footnotesize{}9.95} & {\footnotesize{}10.01} & {\footnotesize{}43} & {\footnotesize{}57}\tabularnewline
 & {\footnotesize{}GL-Uni} & {\footnotesize{}20.18} & {\footnotesize{}21.25} & {\footnotesize{}26.30} & {\footnotesize{}28} & {\footnotesize{}64} & {\footnotesize{}16.53} & {\footnotesize{}19.97} & {\footnotesize{}19.98} & {\footnotesize{}34} & {\footnotesize{}64}\tabularnewline
{\footnotesize{}$\delta^{0}=0.6$} & {\footnotesize{}OLS} & {\footnotesize{}19.80} & {\footnotesize{}24.62} & {\footnotesize{}26.25} & {\footnotesize{}21} & {\footnotesize{}57} & {\footnotesize{}17.34} & {\footnotesize{}22.39} & {\footnotesize{}22.34} & {\footnotesize{}37} & {\footnotesize{}65}\tabularnewline
 & {\footnotesize{}GL-CR} & {\footnotesize{}12.84} & {\footnotesize{}13.66} & {\footnotesize{}18.23} & {\footnotesize{}30} & {\footnotesize{}56} & {\footnotesize{}9.96} & {\footnotesize{}11.93} & {\footnotesize{}11.99} & {\footnotesize{}38} & {\footnotesize{}58}\tabularnewline
 & {\footnotesize{}GL-CR-Iter} & {\footnotesize{}14.85} & {\footnotesize{}11.52} & {\footnotesize{}17.56} & {\footnotesize{}29} & {\footnotesize{}52} & {\footnotesize{}7.26} & {\footnotesize{}9.20} & {\footnotesize{}9.22} & {\footnotesize{}44} & {\footnotesize{}55}\tabularnewline
 & {\footnotesize{}GL-Uni} & {\footnotesize{}16.04} & {\footnotesize{}20.05} & {\footnotesize{}22.77} & {\footnotesize{}26} & {\footnotesize{}56} & {\footnotesize{}13.85} & {\footnotesize{}17.81} & {\footnotesize{}17.94} & {\footnotesize{}38} & {\footnotesize{}60}\tabularnewline
{\footnotesize{}$\delta^{0}=1$} & {\footnotesize{}OLS} & {\footnotesize{}11.69} & {\footnotesize{}18.43} & {\footnotesize{}19.26} & {\footnotesize{}27} & {\footnotesize{}40} & {\footnotesize{}9.38} & {\footnotesize{}14.40} & {\footnotesize{}14.40} & {\footnotesize{}46} & {\footnotesize{}54}\tabularnewline
 & {\footnotesize{}GL-CR} & {\footnotesize{}6.82} & {\footnotesize{}10.85} & {\footnotesize{}12.81} & {\footnotesize{}27} & {\footnotesize{}38} & {\footnotesize{}6.96} & {\footnotesize{}9.43} & {\footnotesize{}9.52} & {\footnotesize{}44} & {\footnotesize{}53}\tabularnewline
 & {\footnotesize{}GL-CR-Iter} & {\footnotesize{}10.67} & {\footnotesize{}7.54} & {\footnotesize{}13.02} & {\footnotesize{}30} & {\footnotesize{}39} & {\footnotesize{}4.44} & {\footnotesize{}6.71} & {\footnotesize{}6.85} & {\footnotesize{}47} & {\footnotesize{}53}\tabularnewline
 & {\footnotesize{}GL-Uni} & {\footnotesize{}9.44} & {\footnotesize{}14.60} & {\footnotesize{}15.15} & {\footnotesize{}27} & {\footnotesize{}37} & {\footnotesize{}8.17} & {\footnotesize{}12.34} & {\footnotesize{}12.34} & {\footnotesize{}45} & {\footnotesize{}54}\tabularnewline
\hline 
\end{tabular}{\footnotesize\par}
\par\end{centering}
\end{singlespace}
\noindent\begin{minipage}[t]{1\columnwidth}%
{\tiny{}The model is $y_{t}=\delta_{1}^{0}+\delta^{0}\mathbf{1}_{\left\{ t>\left\lfloor T\lambda_{0}\right\rfloor \right\} }+e_{t},\,e_{t}=0.6e_{t-1}+u_{t},\,u_{t}\sim i.i.d.\,\mathscr{N}\left(0,\,0.49\right),\,T=100$.
The notes of Table \ref{Table M3 Bias} apply.}%
\end{minipage}
\end{table}

\begin{table}[H]
\caption{\label{Table M5 Bias}Small-sample accuracy of the estimates of the
break point $T_{b}^{0}$ for model M3}

\begin{centering}
{\footnotesize{}}%
\begin{tabular}{ccccccc|ccccc}
\hline 
 &  & {\footnotesize{}MAE} & {\footnotesize{}Std} & {\footnotesize{}$\textrm{RMSE}$} & {\footnotesize{}$Q_{0.25}$} & \multicolumn{1}{c}{{\footnotesize{}$Q_{0.75}$}} & {\footnotesize{}MAE} & {\footnotesize{}Std} & {\footnotesize{}$\textrm{RMSE}$} & {\footnotesize{}$Q_{0.25}$} & {\footnotesize{}$Q_{0.75}$}\tabularnewline
\cline{3-12} \cline{4-12} \cline{5-12} \cline{6-12} \cline{7-12} \cline{8-12} \cline{9-12} \cline{10-12} \cline{11-12} \cline{12-12} 
 &  & \multicolumn{5}{c|}{{\footnotesize{}$\lambda_{0}=0.3$}} & \multicolumn{5}{c}{{\footnotesize{}$\lambda_{0}=0.5$}}\tabularnewline
{\footnotesize{}$\delta^{0}=0.3$} & {\footnotesize{}OLS} & {\footnotesize{}22.33} & {\footnotesize{}26.93} & {\footnotesize{}29.88} & {\footnotesize{}25} & {\footnotesize{}43} & {\footnotesize{}17.23} & {\footnotesize{}22.54} & {\footnotesize{}22.48} & {\footnotesize{}36} & {\footnotesize{}64}\tabularnewline
 & {\footnotesize{}GL-CR} & {\footnotesize{}12.79} & {\footnotesize{}13.32} & {\footnotesize{}18.41} & {\footnotesize{}29} & {\footnotesize{}57} & {\footnotesize{}10.18} & {\footnotesize{}12.11} & {\footnotesize{}12.16} & {\footnotesize{}38} & {\footnotesize{}62}\tabularnewline
 & {\footnotesize{}GL-CR-Iter} & {\footnotesize{}13.92} & {\footnotesize{}15.42} & {\footnotesize{}19.05} & {\footnotesize{}29} & {\footnotesize{}56} & {\footnotesize{}10.85} & {\footnotesize{}13.62} & {\footnotesize{}13.69} & {\footnotesize{}39} & {\footnotesize{}61}\tabularnewline
 & {\footnotesize{}GL-Uni} & {\footnotesize{}15.92} & {\footnotesize{}16.81} & {\footnotesize{}21.40} & {\footnotesize{}30} & {\footnotesize{}55} & {\footnotesize{}11.28} & {\footnotesize{}14.54} & {\footnotesize{}14.54} & {\footnotesize{}41} & {\footnotesize{}59}\tabularnewline
{\footnotesize{}$\delta^{0}=0.4$} & {\footnotesize{}OLS} & {\footnotesize{}22.30} & {\footnotesize{}26.62} & {\footnotesize{}29.55} & {\footnotesize{}22} & {\footnotesize{}61} & {\footnotesize{}13.98} & {\footnotesize{}19.70} & {\footnotesize{}19.68} & {\footnotesize{}42} & {\footnotesize{}58}\tabularnewline
 & {\footnotesize{}GL-CR} & {\footnotesize{}10.08} & {\footnotesize{}12.73} & {\footnotesize{}16.23} & 29 & {\footnotesize{}46} & {\footnotesize{}9.08} & {\footnotesize{}11.41} & {\footnotesize{}11.41} & {\footnotesize{}42} & {\footnotesize{}58}\tabularnewline
 & {\footnotesize{}GL-CR-Iter} & {\footnotesize{}11.08} & {\footnotesize{}14.13} & {\footnotesize{}16.12} & {\footnotesize{}29} & {\footnotesize{}46} & {\footnotesize{}9.10} & {\footnotesize{}12.25} & {\footnotesize{}12.28} & {\footnotesize{}43} & {\footnotesize{}57}\tabularnewline
 & {\footnotesize{}GL-Uni} & {\footnotesize{}12.87} & {\footnotesize{}15.64} & {\footnotesize{}18.53} & {\footnotesize{}29} & {\footnotesize{}49} & {\footnotesize{}9.62} & {\footnotesize{}12.98} & {\footnotesize{}12.97} & {\footnotesize{}43} & {\footnotesize{}57}\tabularnewline
{\footnotesize{}$\delta^{0}=0.6$} & {\footnotesize{}OLS} & {\footnotesize{}8.56} & {\footnotesize{}15.91} & {\footnotesize{}16.15} & {\footnotesize{}27} & {\footnotesize{}33} & {\footnotesize{}8.04} & {\footnotesize{}13.24} & {\footnotesize{}13.22} & {\footnotesize{}46} & {\footnotesize{}54}\tabularnewline
 & {\footnotesize{}GL-CR} & {\footnotesize{}4.68} & {\footnotesize{}8.99} & {\footnotesize{}10.13} & {\footnotesize{}29} & {\footnotesize{}35} & {\footnotesize{}5.62} & {\footnotesize{}8.15} & {\footnotesize{}8.15} & {\footnotesize{}47} & {\footnotesize{}53}\tabularnewline
 & {\footnotesize{}GL-CR-Iter} & {\footnotesize{}7.06} & {\footnotesize{}10.90} & {\footnotesize{}11.35} & {\footnotesize{}29} & {\footnotesize{}35} & {\footnotesize{}6.17} & {\footnotesize{}9.36} & {\footnotesize{}9.36} & {\footnotesize{}46} & {\footnotesize{}53}\tabularnewline
 & {\footnotesize{}GL-Uni} & {\footnotesize{}7.77} & {\footnotesize{}11.85} & {\footnotesize{}12.79} & {\footnotesize{}28} & {\footnotesize{}38} & {\footnotesize{}6.40} & {\footnotesize{}9.63} & {\footnotesize{}9.62} & {\footnotesize{}46} & {\footnotesize{}54}\tabularnewline
{\footnotesize{}$\delta^{0}=1$} & {\footnotesize{}OLS} & {\footnotesize{}2.55} & {\footnotesize{}5.39} & {\footnotesize{}5.45} & {\footnotesize{}29} & {\footnotesize{}31} & {\footnotesize{}2.42} & {\footnotesize{}4.52} & {\footnotesize{}4.52} & {\footnotesize{}49} & {\footnotesize{}51}\tabularnewline
 & {\footnotesize{}GL-CR} & {\footnotesize{}1.24} & {\footnotesize{}3.26} & {\footnotesize{}3.48} & {\footnotesize{}29} & {\footnotesize{}31} & {\footnotesize{}2.31} & {\footnotesize{}4.11} & {\footnotesize{}4.11} & {\footnotesize{}49} & {\footnotesize{}51}\tabularnewline
 & {\footnotesize{}GL-CR-Iter} & {\footnotesize{}2.06} & {\footnotesize{}5.30} & {\footnotesize{}5.33} & {\footnotesize{}29} & {\footnotesize{}31} & {\footnotesize{}2.33} & {\footnotesize{}4.56} & {\footnotesize{}4.59} & {\footnotesize{}48} & {\footnotesize{}52}\tabularnewline
 & {\footnotesize{}GL-Uni} & {\footnotesize{}3.07} & {\footnotesize{}6.18} & {\footnotesize{}6.21} & {\footnotesize{}29} & {\footnotesize{}31} & {\footnotesize{}2.65} & {\footnotesize{}4.79} & {\footnotesize{}4.81} & {\footnotesize{}48} & {\footnotesize{}51}\tabularnewline
\hline 
\end{tabular}{\footnotesize\par}
\par\end{centering}
\noindent\begin{minipage}[t]{1\columnwidth}%
{\tiny{}The model is $y_{t}=1.4\varrho^{0}\delta^{0}\mathbf{1}_{\left\{ t>\left\lfloor T\lambda_{0}\right\rfloor \right\} }+\varrho^{0}y_{t-1}+e_{t},\,e_{t}\sim i.i.d.\,\mathscr{N}\left(0,\,0.5\right),\,\varrho^{0}=0.6,\,T=100$.
The notes of Table \ref{Table M3 Bias} apply.}%
\end{minipage}
\end{table}

\begin{table}[H]
\caption{\label{Table M10}Small-sample coverage rates and lengths of the confidence
sets for model M1}

\begin{centering}
{\footnotesize{}}%
\begin{tabular}{cccccccccc}
\hline 
 &  & \multicolumn{2}{c}{{\footnotesize{}$\delta^{0}=0.4$}} & \multicolumn{2}{c}{{\footnotesize{}$\delta^{0}=0.8$}} & \multicolumn{2}{c}{{\footnotesize{}$\delta^{0}=1.2$}} & \multicolumn{2}{c}{{\footnotesize{}$\delta^{0}=1.6$}}\tabularnewline
 &  & {\footnotesize{}$\textrm{Cov.}$} & {\footnotesize{}$\textrm{Lgth.}$} & {\footnotesize{}$\textrm{Cov.}$} & {\footnotesize{}$\textrm{Lgth.}$} & {\footnotesize{}$\textrm{Cov.}$} & {\footnotesize{}$\textrm{Lgth.}$} & {\footnotesize{}$\textrm{Cov.}$} & {\footnotesize{}$\textrm{Lgth.}$}\tabularnewline
\cline{3-10} \cline{4-10} \cline{5-10} \cline{6-10} \cline{7-10} \cline{8-10} \cline{9-10} \cline{10-10} 
{\footnotesize{}$\lambda_{0}=0.5$} & {\footnotesize{}OLS-CR} & {\footnotesize{}0.959} & {\footnotesize{}80.98} & {\footnotesize{}0.958} & {\footnotesize{}62.61} & {\footnotesize{}0.958} & {\footnotesize{}33.03} & {\footnotesize{}0.937} & {\footnotesize{}15.60}\tabularnewline
 & {\footnotesize{}Bai (1997)} & {\footnotesize{}0.803} & {\footnotesize{}60.84} & {\footnotesize{}0.837} & {\footnotesize{}32.03} & {\footnotesize{}0.874} & {\footnotesize{}15.84} & {\footnotesize{}0.875} & {\footnotesize{}9.34}\tabularnewline
 & {\footnotesize{}$\widehat{U}_{T}\left(T_{\textrm{m}}\right).\textrm{neq}$} & {\footnotesize{}0.945} & {\footnotesize{}80.55} & {\footnotesize{}0.945} & {\footnotesize{}53.41} & {\footnotesize{}0.945} & {\footnotesize{}31.68} & {\footnotesize{}0.945} & {\footnotesize{}22.05}\tabularnewline
 & {\footnotesize{}ILR} & {\footnotesize{}0.962} & {\footnotesize{}77.99} & {\footnotesize{}0.970} & {\footnotesize{}39.64} & {\footnotesize{}0.977} & {\footnotesize{}18.00} & {\footnotesize{}0.985} & {\footnotesize{}10.41}\tabularnewline
 & {\footnotesize{}GL-CR} & {\footnotesize{}0.924} & {\footnotesize{}61.86} & {\footnotesize{}0.919} & {\footnotesize{}53.51} & {\footnotesize{}0.904} & {\footnotesize{}25.99} & {\footnotesize{}0.877} & {\footnotesize{}18.82}\tabularnewline
 & {\footnotesize{}GL-CR-Iter} & {\footnotesize{}0.957} & {\footnotesize{}81.87} & {\footnotesize{}0.958} & {\footnotesize{}64.07} & {\footnotesize{}0.957} & {\footnotesize{}33.04} & {\footnotesize{}0.938} & {\footnotesize{}15.64}\tabularnewline
 & {\footnotesize{}sup-W} & \multicolumn{2}{c}{{\footnotesize{}0.300}} & \multicolumn{2}{c}{{\footnotesize{}0.865}} & \multicolumn{2}{c}{{\footnotesize{}0.996}} & \multicolumn{2}{c}{{\footnotesize{}0.997}}\tabularnewline
\cline{3-10} \cline{4-10} \cline{5-10} \cline{6-10} \cline{7-10} \cline{8-10} \cline{9-10} \cline{10-10} 
{\footnotesize{}$\lambda_{0}=0.35$} & {\footnotesize{}OLS-CR} & {\footnotesize{}0.960} & {\footnotesize{}79.26} & {\footnotesize{}0.920} & {\footnotesize{}61.18} & {\footnotesize{}0.956} & {\footnotesize{}32.07} & {\footnotesize{}0.938} & {\footnotesize{}15.58}\tabularnewline
 & {\footnotesize{}Bai (1997)} & {\footnotesize{}0.828} & {\footnotesize{}59.47} & {\footnotesize{}0.833} & {\footnotesize{}43.92} & {\footnotesize{}0.871} & {\footnotesize{}15.96} & {\footnotesize{}0.877} & {\footnotesize{}9.39}\tabularnewline
 & {\footnotesize{}$\widehat{U}_{T}\left(T_{\textrm{m}}\right).\textrm{neq}$} & {\footnotesize{}0.949} & {\footnotesize{}81.58} & {\footnotesize{}0.940} & {\footnotesize{}68.51} & {\footnotesize{}0.949} & {\footnotesize{}33.23} & {\footnotesize{}0.949} & {\footnotesize{}22.53}\tabularnewline
 & {\footnotesize{}ILR} & {\footnotesize{}0.963} & {\footnotesize{}79.92} & {\footnotesize{}0.972} & {\footnotesize{}44.71} & {\footnotesize{}0.978} & {\footnotesize{}19.81} & {\footnotesize{}0.980} & {\footnotesize{}11.01}\tabularnewline
 & {\footnotesize{}GL-CR} & {\footnotesize{}0.929} & {\footnotesize{}61.18} & {\footnotesize{}0.918} & {\footnotesize{}50.61} & {\footnotesize{}0.900} & {\footnotesize{}32.84} & {\footnotesize{}0.871} & {\footnotesize{}17.96}\tabularnewline
 & {\footnotesize{}GL-CR-Iter} & {\footnotesize{}0.965} & {\footnotesize{}80.58} & {\footnotesize{}0.927} & {\footnotesize{}63.34} & {\footnotesize{}0.957} & {\footnotesize{}32.64} & {\footnotesize{}0.939} & {\footnotesize{}15.65}\tabularnewline
 & {\footnotesize{}sup-W} & \multicolumn{2}{c}{{\footnotesize{}0.276}} & \multicolumn{2}{c}{{\footnotesize{}0.592}} & \multicolumn{2}{c}{{\footnotesize{}0.992}} & \multicolumn{2}{c}{{\footnotesize{}1.000}}\tabularnewline
\cline{3-10} \cline{4-10} \cline{5-10} \cline{6-10} \cline{7-10} \cline{8-10} \cline{9-10} \cline{10-10} 
{\footnotesize{}$\lambda_{0}=0.2$} & {\footnotesize{}OLS-CR} & {\footnotesize{}0.954} & {\footnotesize{}78.14} & {\footnotesize{}0.967} & {\footnotesize{}54.93} & {\footnotesize{}0.972} & {\footnotesize{}29.74} & {\footnotesize{}0.951} & {\footnotesize{}15.73}\tabularnewline
 & {\footnotesize{}Bai (1997)} & {\footnotesize{}0.827} & {\footnotesize{}59.48} & {\footnotesize{}0.884} & {\footnotesize{}33.24} & {\footnotesize{}0.900} & {\footnotesize{}17.11} & {\footnotesize{}0.896} & {\footnotesize{}9.84}\tabularnewline
 & {\footnotesize{}$\widehat{U}_{T}\left(T_{\textrm{m}}\right).\textrm{neq}$} & {\footnotesize{}0.951} & {\footnotesize{}84.04} & {\footnotesize{}0.951} & {\footnotesize{}65.09} & {\footnotesize{}0.951} & {\footnotesize{}57.96} & {\footnotesize{}0.951} & {\footnotesize{}28.16}\tabularnewline
 & {\footnotesize{}ILR} & {\footnotesize{}0.958} & {\footnotesize{}83.06} & {\footnotesize{}0.968} & {\footnotesize{}53.62} & {\footnotesize{}0.978} & {\footnotesize{}25.73} & {\footnotesize{}0.986} & {\footnotesize{}12.69}\tabularnewline
 & {\footnotesize{}GL-CR} & {\footnotesize{}0.914} & {\footnotesize{}60.21} & {\footnotesize{}0.939} & {\footnotesize{}47.40} & {\footnotesize{}0.931} & {\footnotesize{}28.07} & {\footnotesize{}0.910} & {\footnotesize{}14.47}\tabularnewline
 & {\footnotesize{}GL-CR-Iter} & {\footnotesize{}0.965} & {\footnotesize{}79.51} & {\footnotesize{}0.972} & {\footnotesize{}57.07} & {\footnotesize{}0.974} & {\footnotesize{}30.64} & {\footnotesize{}0.957} & {\footnotesize{}15.92}\tabularnewline
 & {\footnotesize{}sup-W} & \multicolumn{2}{c}{{\footnotesize{}0.161}} & \multicolumn{2}{c}{{\footnotesize{}0.576}} & \multicolumn{2}{c}{{\footnotesize{}0.901}} & \multicolumn{2}{c}{{\footnotesize{}0.987}}\tabularnewline
\hline 
\end{tabular}{\footnotesize\par}
\par\end{centering}
\noindent\begin{minipage}[t]{1\columnwidth}%
{\scriptsize{}The model is $y_{t}=\varrho^{0}+Z_{t}\delta_{1}^{0}+Z_{t}\delta^{0}\mathbf{1}_{\left\{ t>\left\lfloor T\lambda_{0}\right\rfloor \right\} }+e_{t},\,Z_{t}=0.3Z_{t-1}+u_{t},\,u_{t}\sim i.i.d.\,\mathscr{N}\left(0,\,1\right)\,e_{t}\sim i.i.d.\,\mathscr{N}\left(0,\,1.21\right),\,T=100$.
Cov. and Lgth. refer to the coverage probability and the average length
of the confidence set (i.e., the average number of dates in the confidence
set). sup-W refers to the rejection probability of the sup-Wald test
using a 5\% asymptotic critical value. The number of simulations is
5,000.}%
\end{minipage}
\end{table}

\begin{table}[H]
\caption{\label{Table M3b}Small-sample coverage rates and lengths of the confidence
sets for model M2}

\begin{centering}
{\footnotesize{}}%
\begin{tabular}{cccccccccc}
\hline 
 &  & \multicolumn{2}{c}{{\footnotesize{}$\delta^{0}=0.4$}} & \multicolumn{2}{c}{{\footnotesize{}$\delta^{0}=0.8$}} & \multicolumn{2}{c}{{\footnotesize{}$\delta^{0}=1.2$}} & \multicolumn{2}{c}{{\footnotesize{}$\delta^{0}=1.6$}}\tabularnewline
 &  & {\footnotesize{}$\textrm{Cov.}$} & {\footnotesize{}$\textrm{Lgth.}$} & {\footnotesize{}$\textrm{Cov.}$} & {\footnotesize{}$\textrm{Lgth.}$} & {\footnotesize{}$\textrm{Cov.}$} & {\footnotesize{}$\textrm{Lgth.}$} & {\footnotesize{}$\textrm{Cov.}$} & {\footnotesize{}$\textrm{Lgth.}$}\tabularnewline
\cline{3-10} \cline{4-10} \cline{5-10} \cline{6-10} \cline{7-10} \cline{8-10} \cline{9-10} \cline{10-10} 
{\footnotesize{}$\lambda_{0}=0.5$} & {\footnotesize{}OLS-CR} & {\footnotesize{}0.910} & {\footnotesize{}67.57} & {\footnotesize{}0.911} & {\footnotesize{}68.87} & {\footnotesize{}0.925} & {\footnotesize{}56.25} & {\footnotesize{}0.945} & {\footnotesize{}42.30}\tabularnewline
 & {\footnotesize{}Bai (1997)} & {\footnotesize{}0.808} & {\footnotesize{}67.57} & {\footnotesize{}0.811} & {\footnotesize{}50.22} & {\footnotesize{}0.843} & {\footnotesize{}32.67} & {\footnotesize{}0.894} & {\footnotesize{}20.74}\tabularnewline
 & {\footnotesize{}$\widehat{U}_{T}\left(T_{\textrm{m}}\right).\textrm{neq}$} & {\footnotesize{}0.981} & {\footnotesize{}91.66} & {\footnotesize{}0.973} & {\footnotesize{}87.38} & {\footnotesize{}0.973} & {\footnotesize{}82.24} & {\footnotesize{}0.973} & {\footnotesize{}79.19}\tabularnewline
 & {\footnotesize{}ILR} & {\footnotesize{}0.944} & {\footnotesize{}83.70} & {\footnotesize{}0.946} & {\footnotesize{}67.01} & {\footnotesize{}0.964} & {\footnotesize{}45.75} & {\footnotesize{}0.980} & {\footnotesize{}28.30}\tabularnewline
 & {\footnotesize{}GL-CR} & {\footnotesize{}0.885} & {\footnotesize{}60.05} & {\footnotesize{}0.884} & {\footnotesize{}52.63} & {\footnotesize{}0.898} & {\footnotesize{}43.41} & {\footnotesize{}0.926} & {\footnotesize{}32.61}\tabularnewline
 & {\footnotesize{}GL-CR-Iter} & {\footnotesize{}0.911} & {\footnotesize{}76.72} & {\footnotesize{}0.911} & {\footnotesize{}69.06} & {\footnotesize{}0.923} & {\footnotesize{}56.31} & {\footnotesize{}0.944} & {\footnotesize{}42.20}\tabularnewline
 & {\footnotesize{}sup-W} & \multicolumn{2}{c}{{\footnotesize{}0.333}} & \multicolumn{2}{c}{{\footnotesize{}0.565}} & \multicolumn{2}{c}{{\footnotesize{}0.878}} & \multicolumn{2}{c}{{\footnotesize{}0.926}}\tabularnewline
\cline{3-10} \cline{4-10} \cline{5-10} \cline{6-10} \cline{7-10} \cline{8-10} \cline{9-10} \cline{10-10} 
{\footnotesize{}$\lambda_{0}=0.35$} & {\footnotesize{}OLS-CR} & {\footnotesize{}0.927} & {\footnotesize{}75.58} & {\footnotesize{}0.910} & {\footnotesize{}66.20} & {\footnotesize{}0.921} & {\footnotesize{}52.89} & {\footnotesize{}0.944} & {\footnotesize{}39.15}\tabularnewline
 & {\footnotesize{}Bai (1997)} & {\footnotesize{}0.838} & {\footnotesize{}66.86} & {\footnotesize{}0.821} & {\footnotesize{}49.34} & {\footnotesize{}0.857} & {\footnotesize{}32.43} & {\footnotesize{}0.893} & {\footnotesize{}20.77}\tabularnewline
 & {\footnotesize{}$\widehat{U}_{T}\left(T_{\textrm{m}}\right).\textrm{neq}$} & {\footnotesize{}0.976} & {\footnotesize{}91.72} & {\footnotesize{}0.976} & {\footnotesize{}87.25} & {\footnotesize{}0.976} & {\footnotesize{}81.76} & {\footnotesize{}0.976} & {\footnotesize{}77.62}\tabularnewline
 & {\footnotesize{}ILR} & {\footnotesize{}0.941} & {\footnotesize{}83.24} & {\footnotesize{}0.945} & {\footnotesize{}68.31} & {\footnotesize{}0.959} & {\footnotesize{}47.58} & {\footnotesize{}0.975} & {\footnotesize{}29.32}\tabularnewline
 & {\footnotesize{}GL-CR} & {\footnotesize{}0.898} & {\footnotesize{}57.32} & {\footnotesize{}0.888} & {\footnotesize{}50.29} & {\footnotesize{}0.902} & {\footnotesize{}43.08} & {\footnotesize{}0.924} & {\footnotesize{}29.06}\tabularnewline
 & {\footnotesize{}GL-CR-Iter} & {\footnotesize{}0.930} & {\footnotesize{}75.87} & {\footnotesize{}0.913} & {\footnotesize{}66.13} & {\footnotesize{}0.921} & {\footnotesize{}52.66} & {\footnotesize{}0.944} & {\footnotesize{}38.71}\tabularnewline
 & {\footnotesize{}sup-W} & \multicolumn{2}{c}{{\footnotesize{}0.489}} & \multicolumn{2}{c}{{\footnotesize{}0.625}} & \multicolumn{2}{c}{{\footnotesize{}0.868}} & \multicolumn{2}{c}{{\footnotesize{}1.000}}\tabularnewline
\cline{3-10} \cline{4-10} \cline{5-10} \cline{6-10} \cline{7-10} \cline{8-10} \cline{9-10} \cline{10-10} 
{\footnotesize{}$\lambda_{0}=0.2$} & {\footnotesize{}OLS-CR} & {\footnotesize{}0.910} & {\footnotesize{}75.24} & {\footnotesize{}0.917} & {\footnotesize{}64.17} & {\footnotesize{}0.931} & {\footnotesize{}48.76} & {\footnotesize{}0.953} & {\footnotesize{}34.26}\tabularnewline
 & {\footnotesize{}Bai (1997)} & {\footnotesize{}0.808} & {\footnotesize{}67.03} & {\footnotesize{}0.852} & {\footnotesize{}50.40} & {\footnotesize{}0.897} & {\footnotesize{}33.62} & {\footnotesize{}0.937} & {\footnotesize{}21.76}\tabularnewline
 & {\footnotesize{}$\widehat{U}_{T}\left(T_{\textrm{m}}\right).\textrm{neq}$} & {\footnotesize{}0.981} & {\footnotesize{}92.17} & {\footnotesize{}0.981} & {\footnotesize{}89.14} & {\footnotesize{}0.981} & {\footnotesize{}84.72} & {\footnotesize{}0.981} & {\footnotesize{}80.39}\tabularnewline
 & {\footnotesize{}ILR} & {\footnotesize{}0.938} & {\footnotesize{}85.37} & {\footnotesize{}0.951} & {\footnotesize{}74.32} & {\footnotesize{}0.963} & {\footnotesize{}57.03} & {\footnotesize{}0.977} & {\footnotesize{}36.24}\tabularnewline
 & {\footnotesize{}GL-CR} & {\footnotesize{}0.912} & {\footnotesize{}56.87} & {\footnotesize{}0.909} & {\footnotesize{}48.68} & {\footnotesize{}0.920} & {\footnotesize{}38.47} & {\footnotesize{}0.932} & {\footnotesize{}23.91}\tabularnewline
 & {\footnotesize{}GL-CR-Iter} & {\footnotesize{}0.894} & {\footnotesize{}75.15} & {\footnotesize{}0.923} & {\footnotesize{}64.14} & {\footnotesize{}0.934} & {\footnotesize{}48.75} & {\footnotesize{}0.953} & {\footnotesize{}34.06}\tabularnewline
 & {\footnotesize{}sup-W} & \multicolumn{2}{c}{{\footnotesize{}0.331}} & \multicolumn{2}{c}{{\footnotesize{}0.565}} & \multicolumn{2}{c}{{\footnotesize{}0.795}} & \multicolumn{2}{c}{{\footnotesize{}0.931}}\tabularnewline
\hline 
\end{tabular}{\footnotesize\par}
\par\end{centering}
\noindent\begin{minipage}[t]{1\columnwidth}%
{\scriptsize{}The model is $y_{t}=\delta_{1}^{0}+\delta^{0}\mathbf{1}_{\left\{ t>\left\lfloor T\lambda_{0}\right\rfloor \right\} }+e_{t},\,e_{t}=0.6e_{t-1}+u_{t},\,u_{t}\sim i.i.d.\,\mathscr{N}\left(0,\,0.49\right),\,T=100$.
The notes of Table \ref{Table M10} apply.}%
\end{minipage}
\end{table}

\begin{table}[H]
\caption{\label{Table M6}Small-sample coverage rates and lengths of the confidence
sets for model M3}

\begin{centering}
{\footnotesize{}}%
\begin{tabular}{cccccccccc}
\hline 
 &  & \multicolumn{2}{c}{{\footnotesize{}$\delta^{0}=0.4$}} & \multicolumn{2}{c}{{\footnotesize{}$\delta^{0}=0.8$}} & \multicolumn{2}{c}{{\footnotesize{}$\delta^{0}=1.2$}} & \multicolumn{2}{c}{{\footnotesize{}$\delta^{0}=1.6$}}\tabularnewline
 &  & {\footnotesize{}$\textrm{Cov.}$} & {\footnotesize{}$\textrm{Lgth.}$} & {\footnotesize{}$\textrm{Cov.}$} & {\footnotesize{}$\textrm{Lgth.}$} & {\footnotesize{}$\textrm{Cov.}$} & {\footnotesize{}$\textrm{Lgth.}$} & {\footnotesize{}$\textrm{Cov.}$} & {\footnotesize{}$\textrm{Lgth.}$}\tabularnewline
\cline{3-10} \cline{4-10} \cline{5-10} \cline{6-10} \cline{7-10} \cline{8-10} \cline{9-10} \cline{10-10} 
{\footnotesize{}$\lambda_{0}=0.5$} & {\footnotesize{}OLS-CR} & {\footnotesize{}0.895} & {\footnotesize{}67.38} & {\footnotesize{}0.923} & {\footnotesize{}39.66} & {\footnotesize{}0.952} & {\footnotesize{}19.23} & {\footnotesize{}0.972} & {\footnotesize{}10.30}\tabularnewline
 & {\footnotesize{}Bai (1997)} & {\footnotesize{}0.759} & {\footnotesize{}43.84} & {\footnotesize{}0.851} & {\footnotesize{}17.50} & {\footnotesize{}0.904} & {\footnotesize{}8.77} & {\footnotesize{}0.935} & {\footnotesize{}5.49}\tabularnewline
 & {\footnotesize{}$\widehat{U}_{T}\left(T_{\textrm{m}}\right).\textrm{neq}$} & {\footnotesize{}0.948} & {\footnotesize{}81.26} & {\footnotesize{}0.947} & {\footnotesize{}71.72} & {\footnotesize{}0.949} & {\footnotesize{}80.79} & {\footnotesize{}0.952} & {\footnotesize{}90.23}\tabularnewline
 & {\footnotesize{}ILR} & {\footnotesize{}0.934} & {\footnotesize{}77.87} & {\footnotesize{}0.951} & {\footnotesize{}50.62} & {\footnotesize{}0.961} & {\footnotesize{}31.01} & {\footnotesize{}0.975} & {\footnotesize{}22.80}\tabularnewline
 & {\footnotesize{}GL-CR} & {\footnotesize{}0.871} & {\footnotesize{}66.58} & {\footnotesize{}0.876} & {\footnotesize{}36.80} & {\footnotesize{}0.893} & {\footnotesize{}15.76} & {\footnotesize{}0.914} & {\footnotesize{}7.50}\tabularnewline
 & {\footnotesize{}GL-CR-Iter} & {\footnotesize{}0.896} & {\footnotesize{}71.50} & {\footnotesize{}0.916} & {\footnotesize{}42.49} & {\footnotesize{}0.948} & {\footnotesize{}20.23} & {\footnotesize{}0.967} & {\footnotesize{}10.30}\tabularnewline
 & {\footnotesize{}sup-W} & \multicolumn{2}{c}{{\footnotesize{}0.300}} & \multicolumn{2}{c}{{\footnotesize{}0.988}} & \multicolumn{2}{c}{{\footnotesize{}0.996}} & \multicolumn{2}{c}{{\footnotesize{}1.000}}\tabularnewline
\cline{3-10} \cline{4-10} \cline{5-10} \cline{6-10} \cline{7-10} \cline{8-10} \cline{9-10} \cline{10-10} 
{\footnotesize{}$\lambda_{0}=0.35$} & {\footnotesize{}OLS-CR} & {\footnotesize{}0.902} & {\footnotesize{}66.58} & {\footnotesize{}0.938} & {\footnotesize{}39.84} & {\footnotesize{}0.953} & {\footnotesize{}19.71} & {\footnotesize{}0.973} & {\footnotesize{}10.47}\tabularnewline
 & {\footnotesize{}Bai (1997)} & {\footnotesize{}0.766} & {\footnotesize{}43.21} & {\footnotesize{}0.855} & {\footnotesize{}17.61} & {\footnotesize{}0.901} & {\footnotesize{}8.82} & {\footnotesize{}0.935} & {\footnotesize{}5.52}\tabularnewline
 & {\footnotesize{}$\widehat{U}_{T}\left(T_{\textrm{m}}\right).\textrm{neq}$} & {\footnotesize{}0.950} & {\footnotesize{}82.13} & {\footnotesize{}0.950} & {\footnotesize{}72.22} & {\footnotesize{}0.952} & {\footnotesize{}78.81} & {\footnotesize{}0.953} & {\footnotesize{}87.02}\tabularnewline
 & {\footnotesize{}ILR} & {\footnotesize{}0.936} & {\footnotesize{}78.13} & {\footnotesize{}0.944} & {\footnotesize{}47.77} & {\footnotesize{}0.959} & {\footnotesize{}30.81} & {\footnotesize{}0.976} & {\footnotesize{}22.16}\tabularnewline
 & {\footnotesize{}GL-CR} & {\footnotesize{}0.885} & {\footnotesize{}66.06} & {\footnotesize{}0.887} & {\footnotesize{}35.98} & {\footnotesize{}0.893} & {\footnotesize{}15.79} & {\footnotesize{}0.917} & {\footnotesize{}7.59}\tabularnewline
 & {\footnotesize{}GL-CR-Iter} & {\footnotesize{}0.908} & {\footnotesize{}71.08} & {\footnotesize{}0.937} & {\footnotesize{}42.99} & {\footnotesize{}0.951} & {\footnotesize{}20.71} & {\footnotesize{}0.967} & {\footnotesize{}10.55}\tabularnewline
 & {\footnotesize{}sup-W} & \multicolumn{2}{c}{{\footnotesize{}0.595}} & \multicolumn{2}{c}{{\footnotesize{}0.973}} & \multicolumn{2}{c}{{\footnotesize{}1.000}} & \multicolumn{2}{c}{{\footnotesize{}1.000}}\tabularnewline
\cline{3-10} \cline{4-10} \cline{5-10} \cline{6-10} \cline{7-10} \cline{8-10} \cline{9-10} \cline{10-10} 
{\footnotesize{}$\lambda_{0}=0.2$} & {\footnotesize{}OLS-CR} & {\footnotesize{}0.910} & {\footnotesize{}65.50} & {\footnotesize{}0.950} & {\footnotesize{}38.42} & {\footnotesize{}0.956} & {\footnotesize{}18.13} & {\footnotesize{}0.969} & {\footnotesize{}10.41}\tabularnewline
 & {\footnotesize{}Bai (1997)} & {\footnotesize{}0.802} & {\footnotesize{}44.43} & {\footnotesize{}0.884} & {\footnotesize{}18.47} & {\footnotesize{}0.952} & {\footnotesize{}8.68} & {\footnotesize{}0.932} & {\footnotesize{}5.32}\tabularnewline
 & {\footnotesize{}$\widehat{U}_{T}\left(T_{\textrm{m}}\right).\textrm{neq}$} & {\footnotesize{}0.948} & {\footnotesize{}89.91} & {\footnotesize{}0.949} & {\footnotesize{}77.46} & {\footnotesize{}0.950} & {\footnotesize{}79.01} & {\footnotesize{}0.932} & {\footnotesize{}88.41}\tabularnewline
 & {\footnotesize{}ILR} & {\footnotesize{}0.932} & {\footnotesize{}79.97} & {\footnotesize{}0.941} & {\footnotesize{}49.72} & {\footnotesize{}0.965} & {\footnotesize{}33.97} & {\footnotesize{}0.951} & {\footnotesize{}21.62}\tabularnewline
 & {\footnotesize{}GL-CR} & {\footnotesize{}0.897} & {\footnotesize{}57.65} & {\footnotesize{}0.903} & {\footnotesize{}34.51} & {\footnotesize{}0.889} & {\footnotesize{}15.21} & {\footnotesize{}0.912} & {\footnotesize{}7.43}\tabularnewline
 & {\footnotesize{}GL-CR-Iter} & {\footnotesize{}0.903} & {\footnotesize{}65.07} & {\footnotesize{}0.923} & {\footnotesize{}37.65} & {\footnotesize{}0.939} & {\footnotesize{}17.46} & {\footnotesize{}0.959} & {\footnotesize{}10.39}\tabularnewline
 & {\footnotesize{}sup-W} & \multicolumn{2}{c}{{\footnotesize{}0.443}} & \multicolumn{2}{c}{{\footnotesize{}0.914}} & \multicolumn{2}{c}{{\footnotesize{}0.999}} & \multicolumn{2}{c}{{\footnotesize{}1.000}}\tabularnewline
\hline 
\end{tabular}{\footnotesize\par}
\par\end{centering}
\noindent\begin{minipage}[t]{1\columnwidth}%
{\scriptsize{}The model is $y_{t}=1.4\varrho^{0}\delta^{0}\mathbf{1}_{\left\{ t>\left\lfloor T\lambda_{0}\right\rfloor \right\} }+\varrho^{0}y_{t-1}+e_{t},\,e_{t}\sim i.i.d.\,\mathscr{N}\left(0,\,0.5\right),\,\varrho^{0}=0.6,\,T=100$.
The notes of Table \ref{Table M10} apply.}%
\end{minipage}
\end{table}

\end{document}